\documentclass[utf8]{FrontiersinVancouver} 

\usepackage{url,hyperref,lineno,microtype,subcaption}
\usepackage[onehalfspacing]{setspace}

\usepackage[]{hyperref} 
\usepackage{hypcap}
\usepackage{xcolor}
\usepackage{caption}
\usepackage{subcaption}
\usepackage{tikz}
\usepackage{graphicx}
\usepackage{algorithm}
\usepackage{algpseudocode}
\usepackage{svg}
\usepackage{booktabs}
\usepackage{threeparttable}
\usepackage{tikz}
\usepackage{tabularx}
\usepackage{makecell}
\usepackage{tablefootnote}
\usepackage{amsmath}
\usepackage{cleveref}
\usepackage{array}
\usepackage{comment}
\usepackage{pdflscape}
\usepackage{color, colortbl}
\usepackage{wrapfig}
\newcolumntype{P}[1]{>{\centering\arraybackslash}p{#1}}
\definecolor{Gray}{gray}{0.9}
\newcolumntype{g}{>{\columncolor{Gray}}r}
\usetikzlibrary{shapes.geometric,arrows,positioning}
\tikzstyle{startstop} = [rectangle, text width=3cm, minimum   height=1cm,text centered, draw=black, fill=red!30]
   \tikzstyle{io} = [trapezium, trapezium left angle=70, trapezium right 
   angle=110, text width=3cm, minimum height=1cm, text centered, draw=black, fill=violet!30]
   \tikzstyle{process} = [rectangle, text width=3cm, minimum height=1cm,    text centered, draw=black, fill=orange!30]
   \tikzstyle{decision} = [diamond, text width=3cm, minimum height=1cm, text centered, draw=black, fill=green!30]
   \tikzstyle{arrow} = [thick,->,>=stealth]

\def\keyFont{\fontsize{8}{11}\helveticabold }
\def\firstAuthorLast{Pinto {et~al.}} 
\def\Authors{Imanol Pinto$^{1,*}$, Álvaro Olazarán$^{1}$, David Jurío$^{1}$, Borja de la Osa$^{1}$, Miguel Sainz$^{1}$, Aritz Oscoz$^{1}$, Jerónimo Ballaz$^{1}$, Javier Gorricho$^{2,3}$, Mikel Galar$^{5,3}$ and José Andonegui$^{3,4}$}
\def\Address{$^{1}$Health Technology Services, General Directorate of Telecommunications and Digitalization (DGTD), C/ Cabarceno 6, Planta 3, Sarriguren 31621, Spain \\
$^{2}$Health Outcomes Evaluation and Dissemination Service, Navarre Public Health Service (SNS-O), Pamplona, Spain  \\
$^{3}$IdiSNA, Navarre Institute of Health Research, C/ Irunlarrea 3, Pamplona 31008, Spain  \\
$^{4}$University Hospital of Navarre, Ophthalmology Service, C/ Irunlarrea 3, Pamplona 31008, Spain \\
$^{5}$Institute of Smart Cities (ISC), Public University of Navarre (UPNA), Campus Arrosadia, 31006, Pamplona, Spain  }
\def\corrAuthor{Corresponding Author}

\def\corrEmail{imanol.pinto.lopez@navarra.es}

\begin{document}
\onecolumn
\firstpage{1}

\title[Improving diabetic retinopathy screening using Artificial Intelligence]{Improving diabetic retinopathy screening using Artificial Intelligence: design, evaluation and before-and-after study of a custom development} 

\author[\firstAuthorLast ]{\Authors} 
\address{} 
\correspondance{} 

\extraAuth{}

\maketitle

\begin{abstract}

\noindent\textbf{Background:} The worst outcomes of diabetic retinopathy (DR) can be prevented by implementing DR screening programs assisted by AI. At the University Hospital of Navarre (HUN), Spain, general practitioners (GPs) grade fundus images in an ongoing DR screening program, referring to a second screening level (ophthalmologist) target patients.

\noindent\textbf{Methods:} After collecting their requirements, HUN decided to develop a custom AI tool, called NaIA-RD, to assist their GPs in DR screening. This paper introduces NaIA-RD, details its implementation, and highlights its unique combination of DR and retinal image quality grading in a single system. Its impact is measured in an unprecedented before-and-after study that compares 19,828 patients screened before NaIA-RD's implementation and 22,962 patients screened after.

\noindent\textbf{Results:} NaIA-RD influenced the screening criteria of 3/4 GPs, increasing their sensitivity.
Agreement between NaIA-RD and the GPs was high for non-referral proposals (94.6\% or more), but lower and variable (from 23.4\% to 86.6\%) for referral proposals. An ophthalmologist discarded a NaIA-RD error in most of contradicted referral proposals by labeling the 93\% of a sample of them as referable. In an autonomous setup, NaIA-RD would have reduced the study visualization workload by 4.27 times without missing a single case of sight-threatening DR referred by a GP.

\noindent\textbf{Conclusion:} DR screening was more effective when supported by NaIA-RD, which could be safely used to autonomously perform the first level of screening. This shows how AI devices, when seamlessly integrated into clinical workflows, can help improve clinical pathways in the long term.

\tiny
 \keyFont{ \section{Keywords:} Diabetic retinopathy, AI medical device, Decision-support system, Deep learning, Before-and-after study} 
\end{abstract}

\section{Introduction}


\label{sec:intro}
Diabetic retinopathy (DR) is the leading cause of vision loss among the working-age population in developed countries \cite{RDPrevalence}, but the worst outcomes can be prevented with early detection and treatment. In fact, the implementation of DR screening programs is recommended by the American Diabetes Association \cite{Standards2003} and the International Council of Ophthalmology \cite{Guidelines2017}. 

DR screening is usually performed by trained personnel who are not necessarily ophthalmologists. They grade (visualize and assess) eye fundus photographs, called retinographies, which are taken with a non-mydriatic digital camera. These graders look for DR signs, such as hemorrhages, and refer (send) the  patient to an ophthalmologist if necessary. Their primary goal is to refer patients who need evaluation by a specialist, who will then decide if treatment or further follow-up is necessary.

Given this workflow, automated methods can bring a more efficient and cost-effective DR screening \cite{DRCostScotland2010, DRCostEngland2016}. Several AI-based medical devices (CE-marked or FDA-approved) are available for this purpose \cite{ReviewAIForDR2019}. These tools promise to eliminate or reduce the burden of manual grading.

However, the performance of AI-based medical devices often degrades when they are used outside the clinical sites from which their data originated \cite{HowMedicalAIAreEvaluated2021, Multicenter2021}. A recent study \cite{Multicenter2021} compared seven algorithms that were being used in clinics, and highlighted the need for prospective, interventional trials for commercialized products, as they measured a wide range of sensitivities (50.98-85.90\%). These interventional studies are not required to obtain a CE mark or FDA approval and are therefore very rare \cite{AIvsClinicians2020}.

To make matters worse, AI-powered medical devices are often negatively affected by their environment: task sharing, user knowledge, infrastructure, integration, and socio-environmental factors are challenges that hinder their success \cite{IntegratingRadiologyFaric2023, GoogleHealth}. This problem is exacerbated when the clinical protocol is already established before it is supported by AI. Sometimes it is simply not feasible to implement a generic AI tool into an ongoing DR screening program.

This is the case at the University Hospital of Navarre (HUN). This hospital in Spain has been offering a public DR screening program since 2015, and has been working with us to support it with AI. We collected HUN's DR screening requirements and found that none of the available CE-marked medical devices could be used without significant limitations and risks. Therefore, we developed a custom, AI-based DR screening tool for HUN: NaIA-RD.

After validating the performance of NaIA-RD using two private and six public datasets \cite{KaggleEyePACSCompetition2016, APTOSKaggle2019, Messidor2018, IDRIDDataset, OIA-DDR-2019, EyeQFu2019}, we deployed it in July 2020, integrated into the Hospital Information System (HIS). It has been used for routine DR screening for more than three years. Using the data from this interventional prospective study, we compared how DR screening was performed before and after the deployment of NaIA-RD, measuring how the tool has influenced clinical decisions.

This paper makes two significant contributions to the literature: First, it measures the impact of an AI tool on real-world clinical decisions. Most published prospective studies typically compare the AI tool's performance to that of manual graders \cite{Abramoff2018USProspectiveIDX, Raumviboonsuk2019ThailandProspective, Gulshan2019IndiaProspective, Ipp2021ProspectiveMulticenterEyeArt, Lim2022ProspectiveMulticenterEyeArt, Heydon2022ProspectiveMulticenterEyeArt, Skevas2022ProspectiveRetCAD, Meredith2023ProspectiveRetCAD}, or they evaluate the tool's ability to reduce the burden of manual grading \cite{Ribeiro2014ProspectiveRetmarker}, but they do not assess how the tool has influenced clinician behavior. Second, this paper details a novel procedure for combining DR grading models with retinal image quality (gradability) models. Our proposed system selects the most clinically suitable image (field of view) and consistently provides independent DR and gradability scores.
We found no other work describing how to integrate both assessments into a single AI system, although there are numerous publications dedicated to each topic individually \cite{DeepLearningDRSurvey2022, GradabilitySurvey2020}.

This paper is organized as follows. First, Section \ref{sec:dr-process} details the DR screening process at HUN, before and after NaIA-RD's assistance, and Section \ref{sec:motivation} summarizes the hospital's requirements and how commercial AI devices do not meet them. Section \ref{sec:development} details NaIA-RD's development, from system design to neural network training, calibration, interpretation, image enhancement, and machine learning operations (MLOps). Then, Section \ref{sec:results} reports the results: First, in laboratory settings (\ref{sec:field-test-set}-\ref{sec:external-validation}) and last, in real clinical settings (\ref{sec:prospective-study}). Section \ref{sec:discussion} discusses NaIA-RD's performance, impact, and limitations. We draw our final conclusions in Section \ref{sec:conclusion}.

\section{Materials and methods}
\subsection{DR Screening at HUN}
\label{sec:dr-process}

In 2015 a DR tele-screening program was set up at HUN. Since then, all patients assigned to the hospital and diagnosed with Type 2 diabetes have been scheduled for annual retinal imaging and screening. Over these years, the number of patients screened has steadily increased, reaching nearly 8,000 in 2023. 

A team of four primary care general practitioners (GPs) --who received specific training \cite{Ando2010, Ando2012}-- have remotely assessed retinal images using a centralized HIS. When they detected signs of referable DR or the eye fundus was non-gradable due to insufficient image quality, they referred (sent) the images (grouped as a study) to a second screening level (an ophthalmologist), who decided whether an on-site eye examination was necessary. 

The following subsections further explain this DR screening protocol (Section \ref{subsec:protocol}) and how it has been supported by AI (Section \ref{subsec:aiassisted}).

\subsubsection{The DR screening protocol} \label{subsec:protocol}
Figure \ref{fig:rd-screening} shows the screening process of HUN using the Business Process Model Notation (BPMN) standard. To model this process, we visited nurses, GPs, and ophthalmologists at the screening sites. Then, we validated our observations using anonymized hospital data. In summary, DR screening at HUN is performed in three main steps:

\begin{figure*}[ht]
    \centering
  \includegraphics[width=0.95\linewidth]{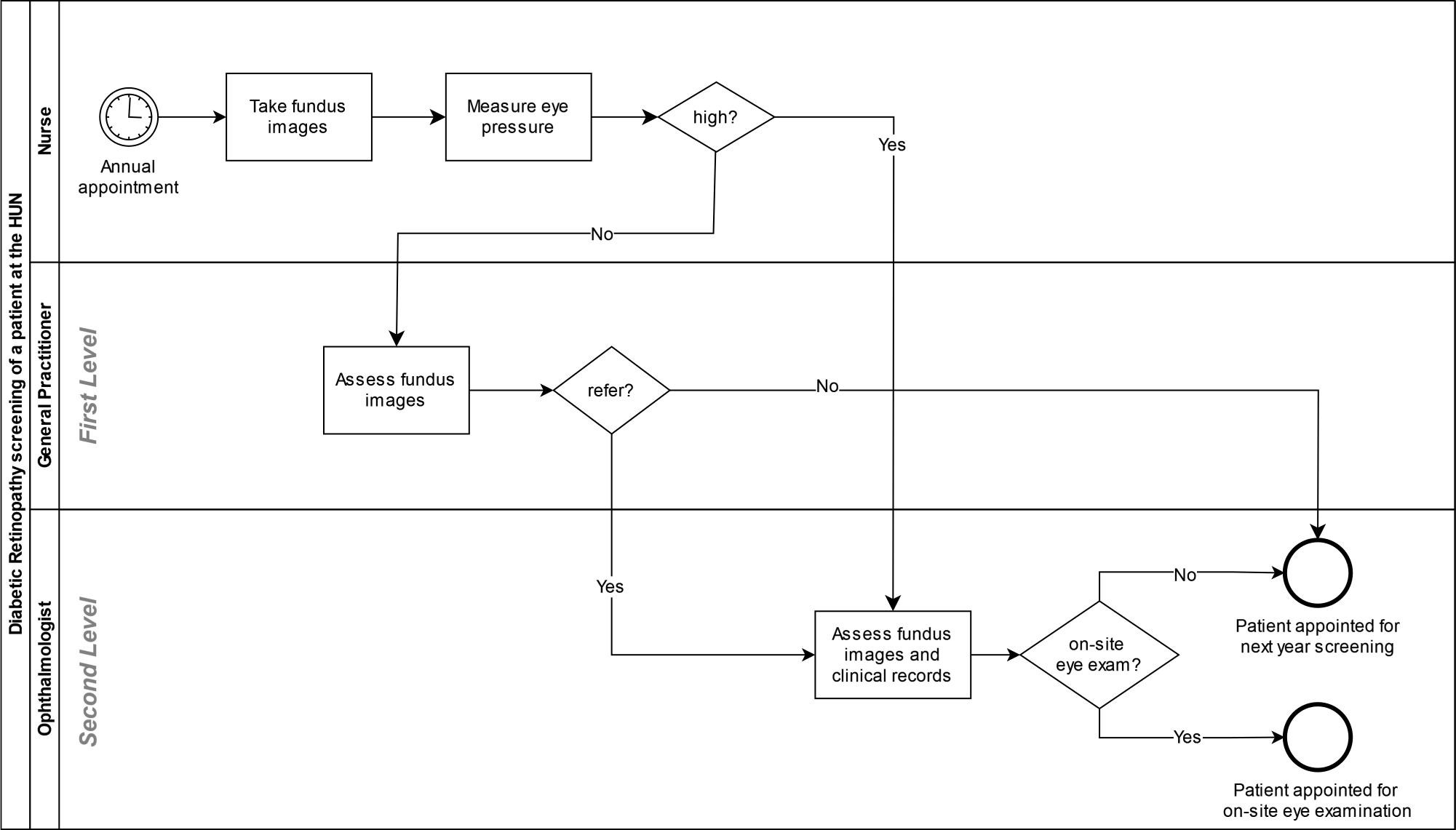}
  \caption{BPMN diagram of the process of screening a patient at HUN. Tasks are represented with squares, diamonds represent bifurcations, and circles represent start and end events. Patients appointed for an on-site eye examination abandon the DR screening program until an ophthalmologist decides otherwise. }
  \label{fig:rd-screening}
\end{figure*}

\begin{enumerate}
    \item \textbf{Image taking (nurse)}. A nurse usually takes two non-mydriatic fundus images: a macula-centered (central) and an optic-disc-centered (nasal) fundus field. However, additional images may be taken if necessary. These fundus images are uploaded to a centralized Picture Acquisition Server (PACS) as a study, which is composed of two eyes (left and right), and will be assessed by a first screening level. The nurse also measures the intraocular pressure\footnote{Intraocular pressure is measured for safety reasons, as ocular hypertension usually does not show any findings in fundus images.}. If it is high, the nurse will refer the study directly to the second screening level.
    
    \item \textbf{First screening level (GP)}. A trained GP visualizes the images, grades DR following the International Clinical Diabetic Retinopathy (ICDR) severity scale \cite{ICDR-2003}, and decides whether the study should be referred to the second screening level. They should refer a study if it is not gradable or if it shows signs of more than mild DR \cite{Guidelines2017}. Their primary goal is to refer patients who need to be scheduled for an on-site eye examination.
    
    \item \textbf{Second screening level (ophthalmologist)}. An ophthalmologist grades the referred study following the ICDR scale \cite{ICDR-2003}\footnote{The ophthalmologist overrides the grade of the GP.} and determines if the patient needs an on-site eye examination. This decision is based on the fundus images, patient history, and glycemic (hemoglobin A1c) measurements. 
\end{enumerate}

\begin{subfigure}
    \centering
    \begin{minipage}[b]{0.8\textwidth}
        \centering
        \includegraphics[width=.20\textwidth]{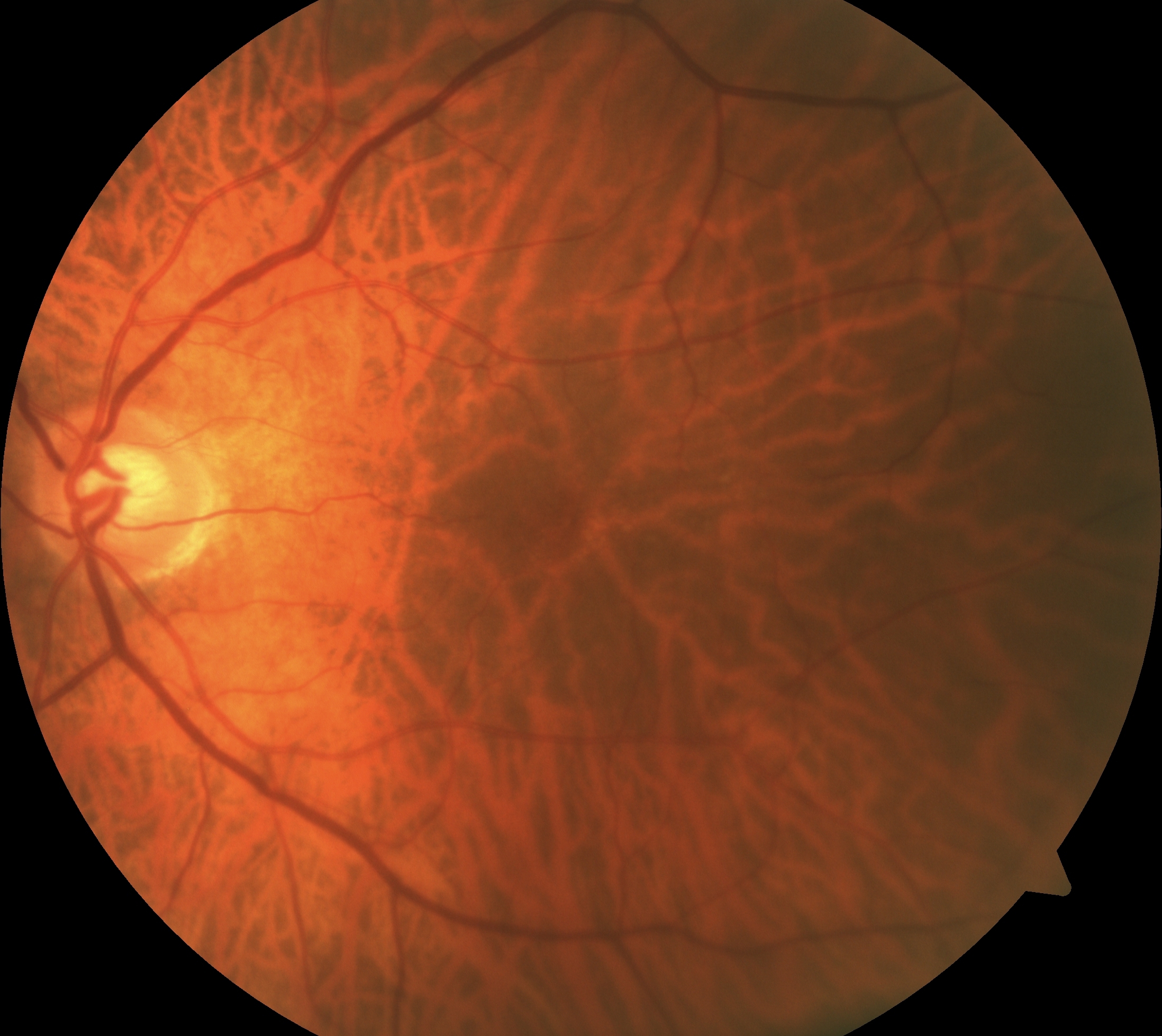}
         \includegraphics[width=.20\textwidth]{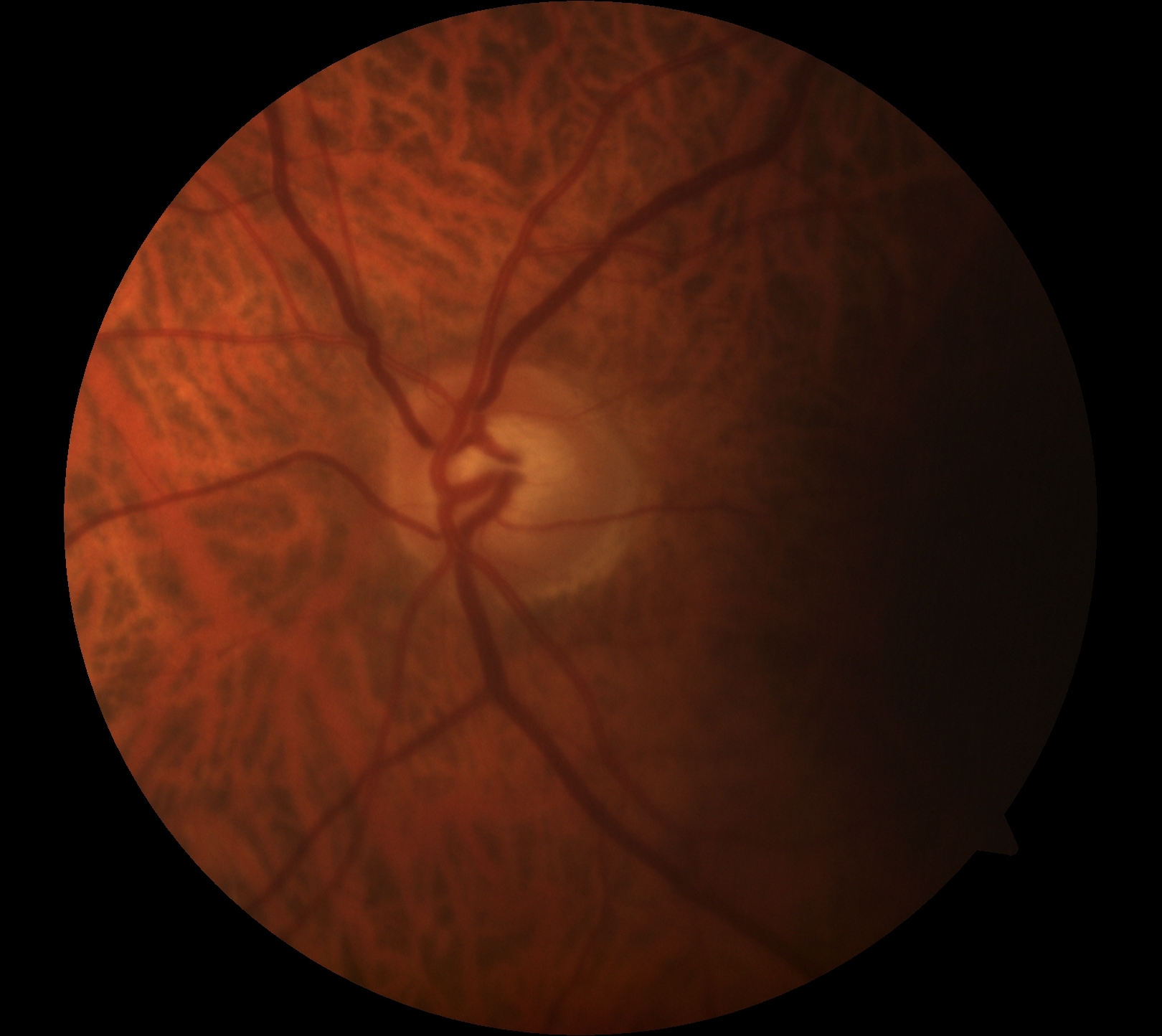}
         \includegraphics[width=.23\textwidth]{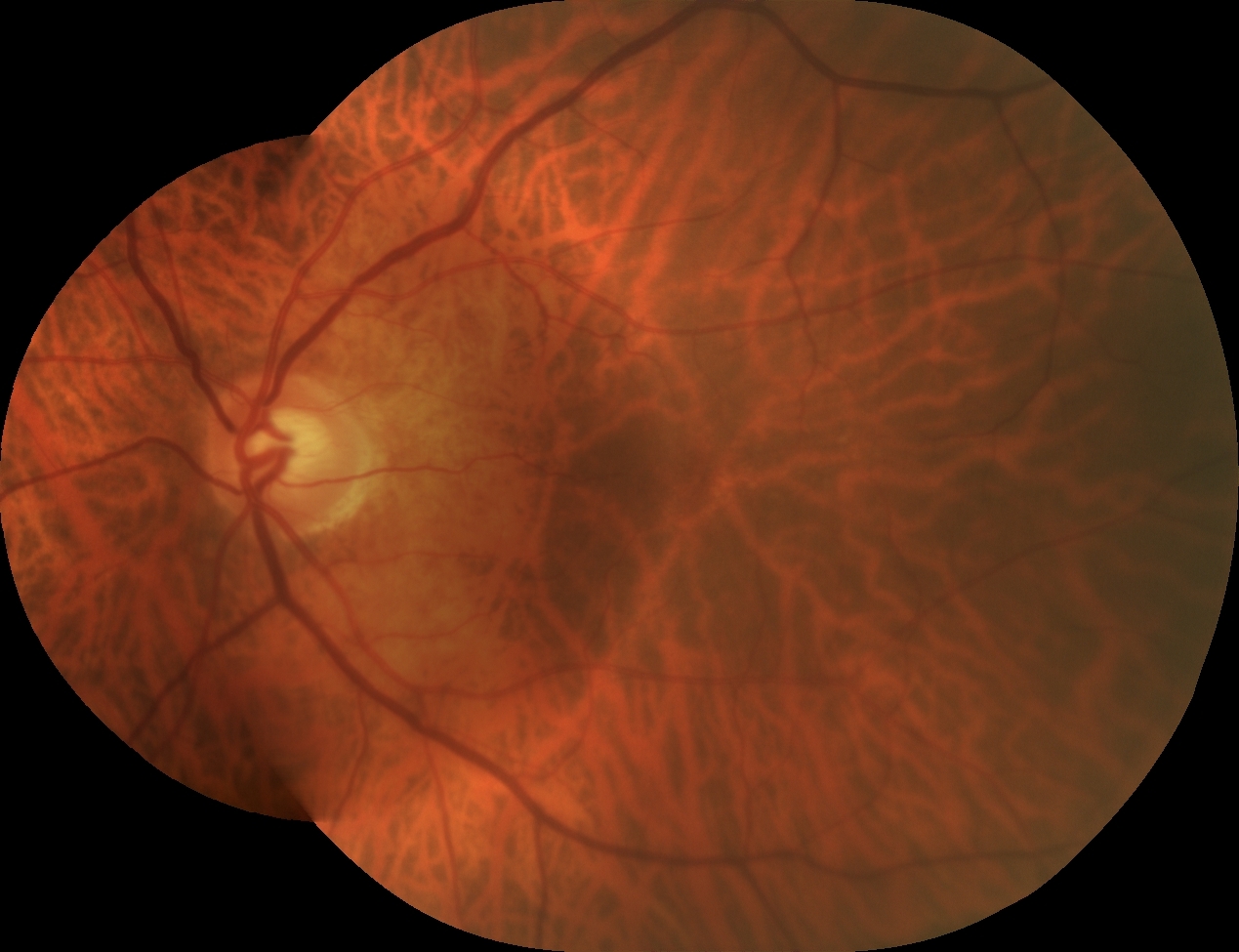}
        \caption{Sample eye study composed by central, nasal and composite fundus fields.}
        \label{fig:study-1}
    \end{minipage}
    \begin{minipage}[b]{0.8\textwidth}
        \vspace*{0.5cm}
        \centering
        \includegraphics[width=.16\textwidth]{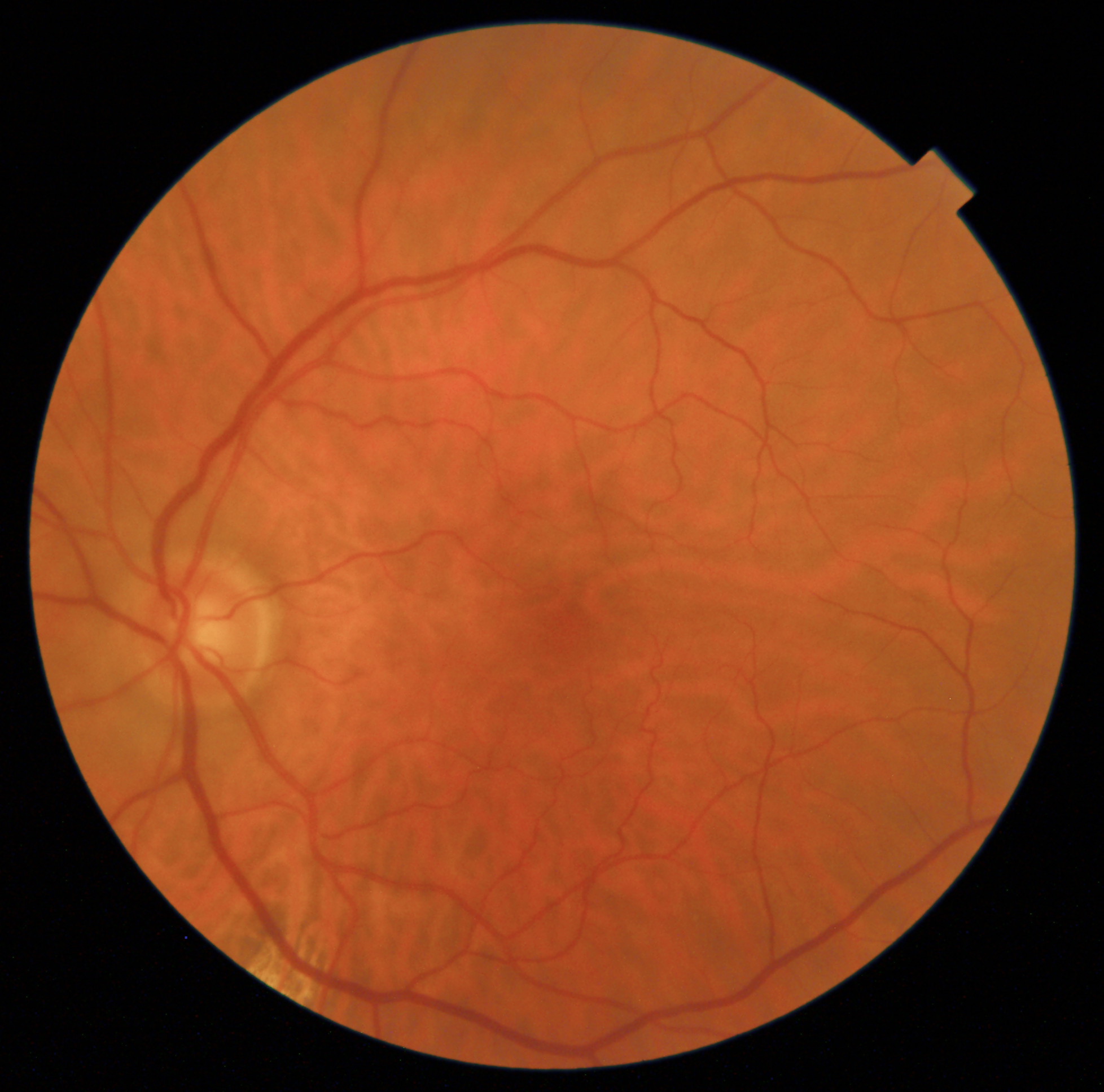}
        \includegraphics[width=.16\textwidth]{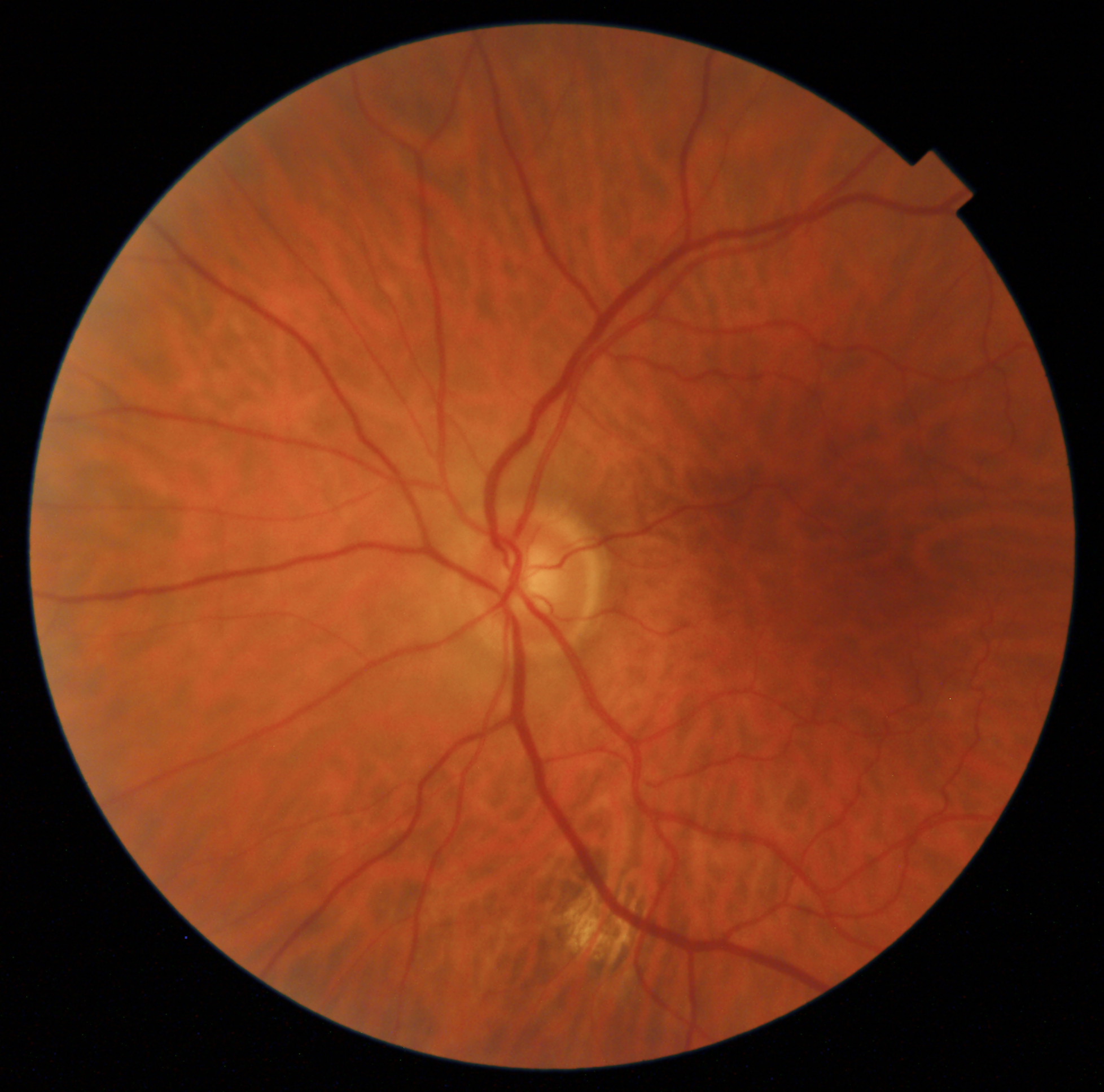}
        \includegraphics[width=.16\textwidth]{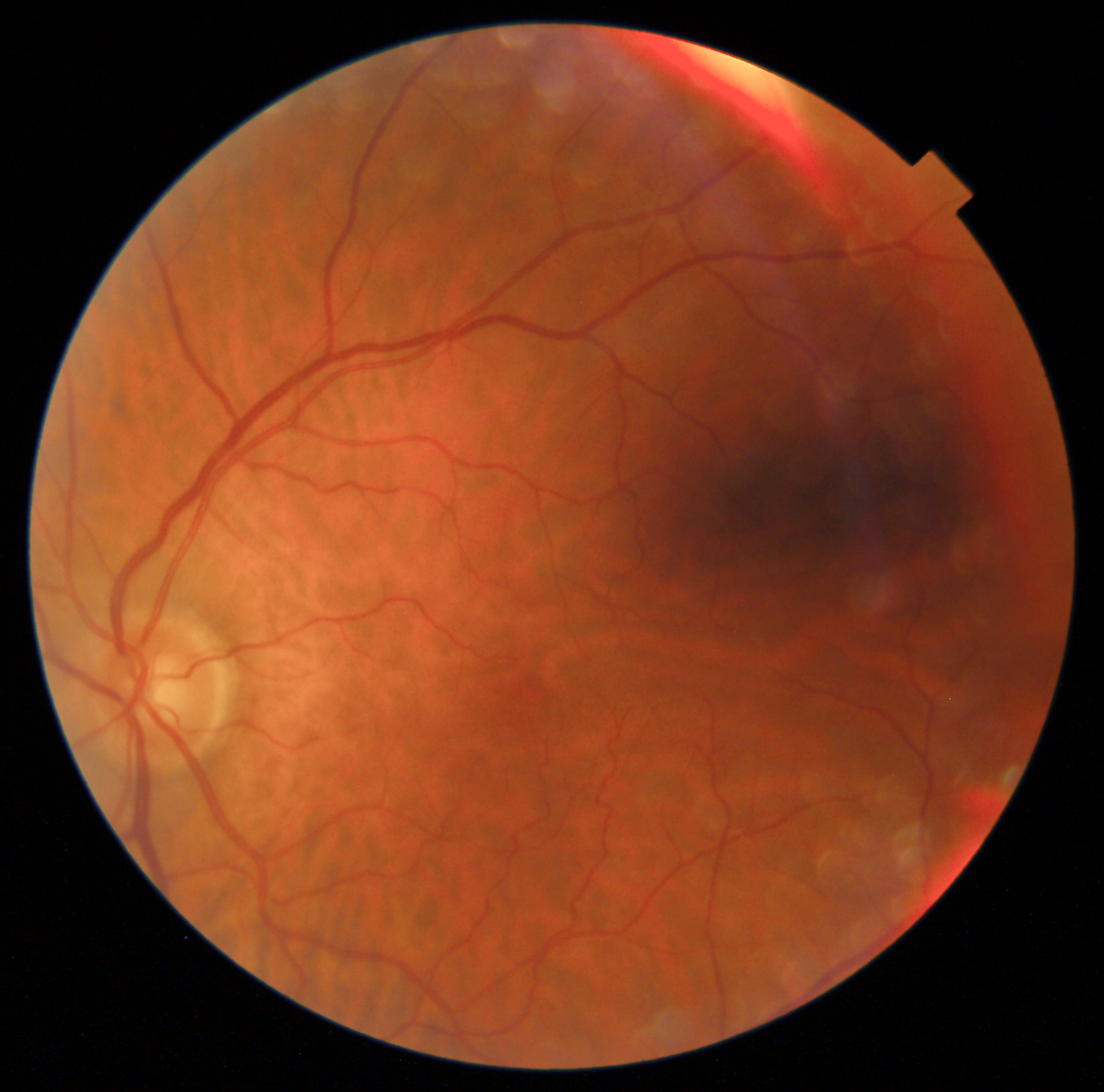}
        \includegraphics[width=.16\textwidth]{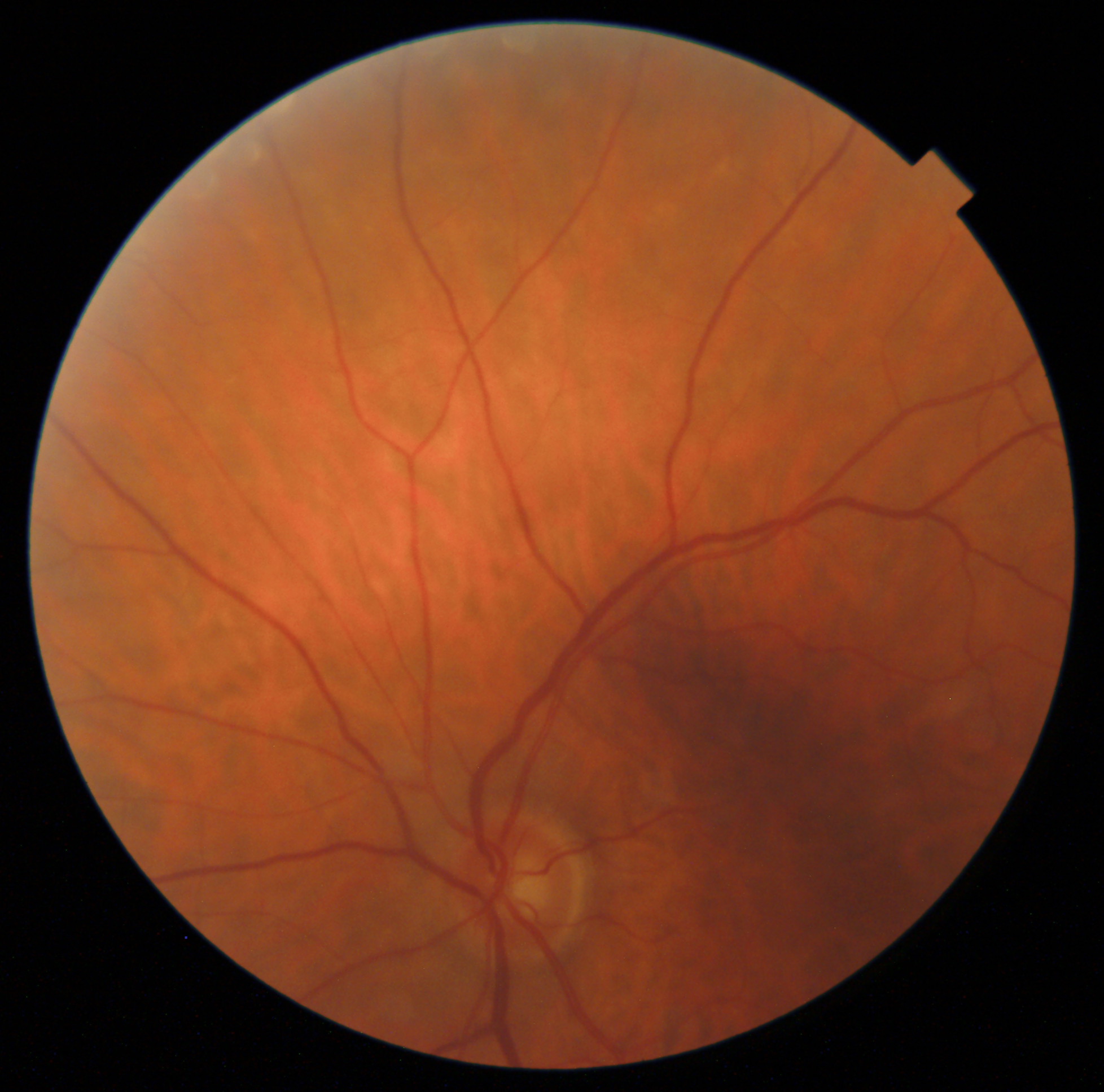}
        \includegraphics[width=.16\textwidth]{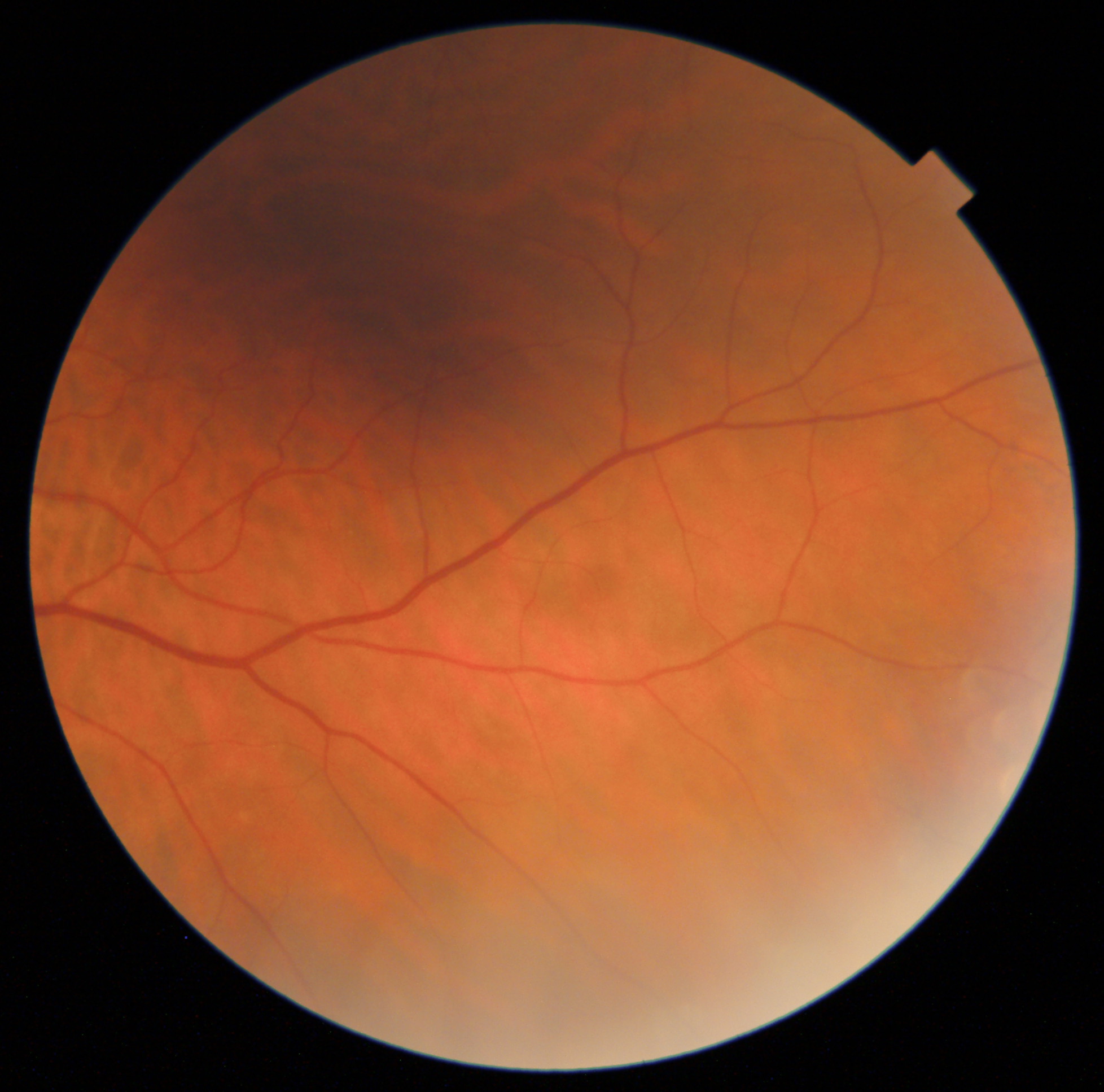}
        \includegraphics[width=.16\textwidth]{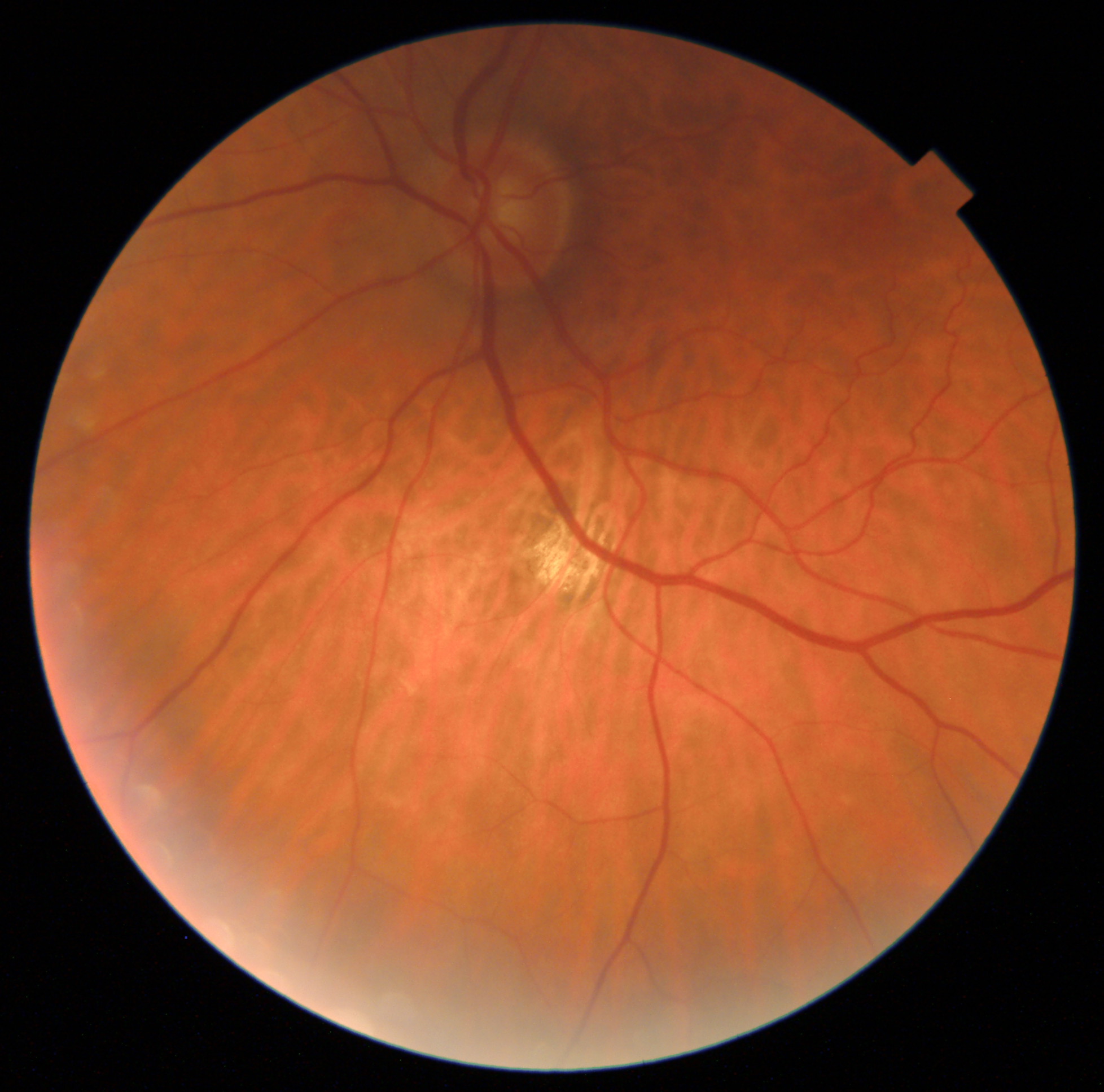}
        \caption{Sample eye study composed by repeated fundus fields: central, nasal, central, OD down, no OD, and OD up (OD refers to Optical Disc).}
        \label{fig:study-2}
    \end{minipage}
    \setcounter{subfigure}{-1}
    \caption{Two sample studies (left eyes only) from the DR screening program of HUN. Note that some fundus fields are often redundant.}
    \label{fig:sample-studies}
\end{subfigure}

If the second screening level determines that an on-site eye examination is needed, the patient will leave the DR screening program (a retina specialist will examine the patient and decide if treatment or other outpatient care is needed). Otherwise, the patient will be scheduled for the next year's DR screening imaging session.

Regarding the imaging protocol, Figure \ref{fig:study-1} shows an example of a screened eye composed of the two non-mydriatic fundus images that the nurse usually takes, which are often accompanied by a composite image. The composite does not add any new clinical information, as it is just a collage of the central and nasal images. However, as we show in Figure \ref{fig:study-2}, there is no guarantee that nurses will strictly follow this protocol. In fact, more images are often included in difficult cases, as it is done in other hospitals \cite{GoogleHealth}.

This screening protocol is based on the guidelines published by the International Council of Ophthalmology in 2017 \cite{Guidelines2017}. Many other DR screening programs around the world follow the same or similar guidelines \cite{DRProgramsOverview2020, DRProgrammeVeterans2008, DRProgrammeScott2015, DRProgrammeSingapore2016, DRProgrammeNHS2017, DRProgrammePortugal2021}, but each program has its own characteristics. For example, the NHS Diabetic Eye Screening Program \cite{DRProgrammeNHS2017} adds an arbitration grading step in case of disagreement between first- and second-level graders and always performs two-field mydriatic photography. Closely related, the Scottish DR Screening Program uses a microaneurysm detection software prior to manual grading and performs single-field non-mydriatic photography \cite{DRProgrammeScott2015}. On the other hand, the Singapore Integrated Diabetic Retinopathy Program centralizes a single human level of screening and is piloting SELENA+ \footnote{SELENA+ official web page: \url{https://www.synapxe.sg/healthtech/health-ai/selena/}}, a Deep Learning system that fully automates the first screening level \cite{ SelenaAppsWee2017, SingaporeSelena2023, SelenaDevelopTing2017}.

\subsubsection{AI assistance}
Since July 2020, the DR screening protocol has been assisted by NaIA-RD. However, the screening protocol has not changed with the AI. In this new setting, GPs review the screening proposal of NaIA-RD before assessing the images, while the rest of the screening steps remain the same. NaIA-RD has enabled the following advanced features:

\begin{itemize}
    \item The HIS shows NaIA-RD's motivated screening proposal for each study. This proposal is a referral (due to DR or non-gradable fundus) or non-referral recommendation. When the proposal is accepted, the clinical report is generated automatically.
    \item The HIS study worklist can be sorted by NaIA-RD's outputs, either by referability (a single probability score) or by category (non-referable, referable DR or non-gradable).
    \item Lesions detected by NaIA-RD are highlighted on the image when the AI recommends a referral due to DR. 
    \item The original fundus image is enhanced by NaIA-RD.
\end{itemize}

\subsection{Motivation for a custom AI development} \label{subsec:aiassisted}
\label{sec:motivation}
When the use of AI to assist in DR screening at HUN was first considered, we gathered the requirements for a tool that could effectively support the process and explored CE-marked products that might meet these needs. However, we found none that fulfilled the criteria. As a result, HUN opted to develop NaIA-RD as an in-house AI solution. This section outlines the reasoning behind this decision. Section \ref{subsec:requirements} describes the requirements, and Section \ref{subsec:commercial} summarizes the commercial medical devices considered.

\subsubsection{Requirements}\label{subsec:requirements}
We met diverse stakeholders from the hospital to collect a broad set of goals, expectations and restrictions: clinicians, ophthalmologists, nurses, IT engineers and managers were interviewed. For brevity, a detailed list of the collected requirements is provided in the Supplementary Material. However, they can be summarized as follows. The AI tool should:

\begin{itemize}
    \item Be compatible with the current DR screening protocol, patient groups and cameras, while allowing for future inclusion of Type I diabetic patients and new camera models.
    \item Assist with the fundus image assessment task of the first screening level, enabling task automation and worklist prioritization in the HIS through data-level integration.
    \item Support workflow orchestration, as well as monitoring of disease prevalence and retinal image quality. 
    \item Offer interpretability and image enhancement features to facilitate human assessment.
\end{itemize}

\subsubsection{Commercial medical devices}\label{subsec:commercial}
Following the requirements, we first explored the acquisition of a commercial solution. We evaluated six Class IIa CE-marked devices: IDx-DR\footnote{IDx-DR: \url{https://www.healthvisors.com/idx-dr/}}, EyeArt\footnote{EyeArt: \url{https://www.eyenuk.com/en/products/eyeart/}}, Retmarker\footnote{Retmarker: \url{https://www.retmarker.com/morescreening/}}, OpthAI\footnote{OpthAI: \url{https://www.ophtai.com/en/}}, RetCad\footnote{RetCad: \url{https://retcad.eu/}} and SELENA+. A detailed comparison of these devices is provided in the Supplementary Material. Note that we discarded products with insufficient public information, those requiring a non-standard 45º field camera, or those self-certified as Class I medical devices.

According to this comparison, we could not find any Class II CE-marked device that met the requirements without significant limitations and risks:

\begin{itemize}
    \item No tool fully supported the current patient population and imaging protocol (including cameras and study format).
    \item Few tools offered a data-level integration API, and those that did, lacked a detailed gradability (image quality) score.
    \item Few tools provided interpretable results or enhanced fundus images.
\end{itemize}

As a consequence, HUN decided to request to the competent governmental units\footnote{The Navarre Public Health Service and the Health Technology Services of the Government of Navarre are the official IT service of HUN.} the development of NaIA-RD, whose details we present in the following sections.

\subsection{Development of a custom AI tool for DR screening}
\label{sec:development}
In this section we describe the technical design of NaIA-RD, including its neural networks, datasets, calibration, interpretability, image enhancement and MLOps. For brevity, a summary of the development project life cycle can be found in the Supplementary Material.

\subsubsection{Architecture}
\label{sec:architecture}
We designed NaIA-RD as a modular system consisting of three neural networks and a software component that orchestrates them. The neural networks are the following: 
\begin{enumerate}
    \item \textit{Field Classifier}: Identifies the field of view of a fundus image  (described in Section \ref{sec:field-selector}).
    \item \textit{Gradability Classifier}: Determines if a fundus image is gradable (described in Section \ref{sec:grad-evaluator}).
    \item \textit{DR Classifier}: Determines if a fundus image shows referable DR (described in Section \ref{sec:dr-evaluator}).
\end{enumerate}

\begin{figure}
\centering
  \includegraphics[width=0.7\linewidth]{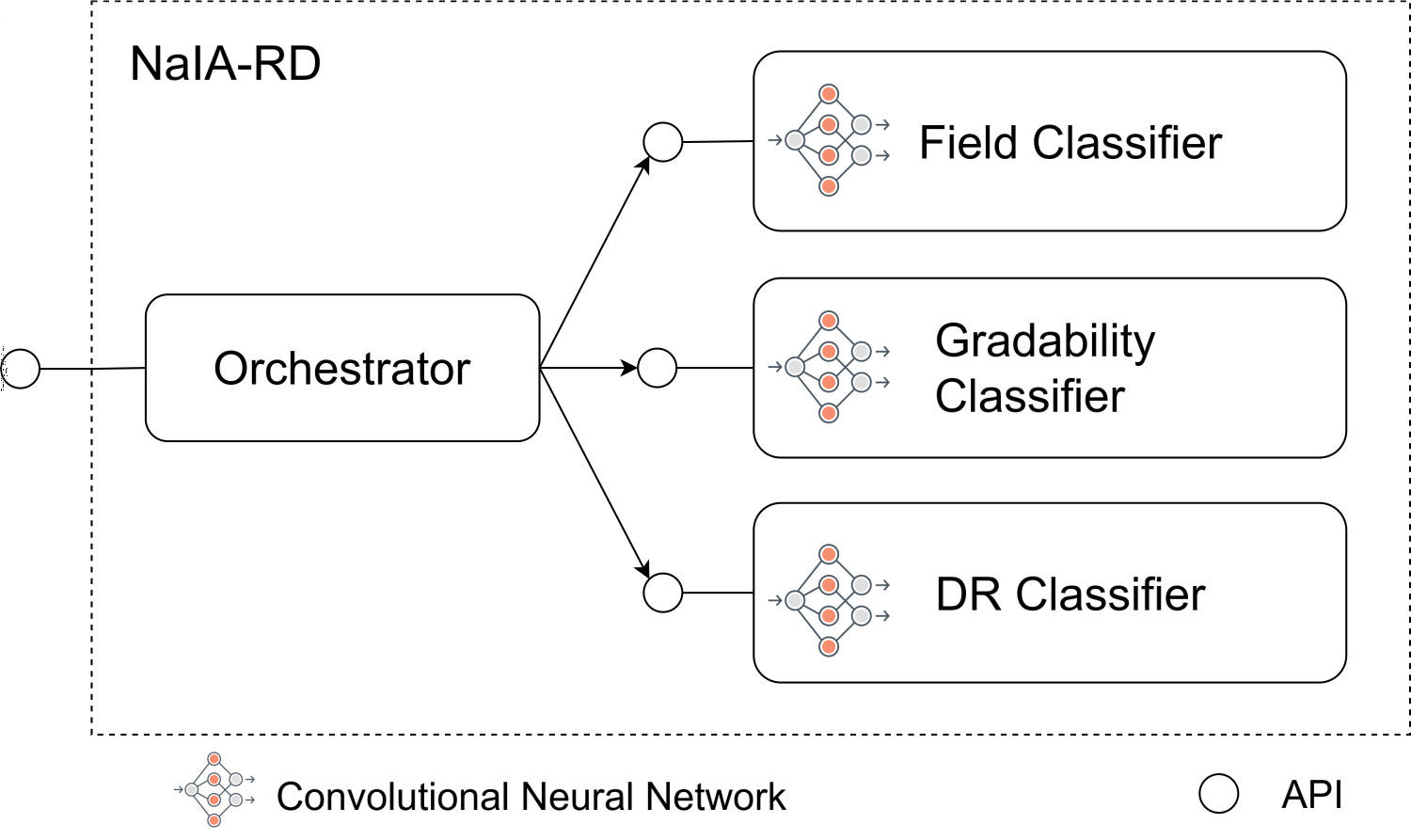}
  \caption{Components of NaIA-RD.}
  \label{fig:components}
\end{figure}

A software component we call Orchestrator uses these neural networks to generate a DR screening proposal in response to an incoming eye screening request from the HIS. The request consists of a list of fundus images belonging to the same eye. Figure \ref{fig:components} summarizes this architecture. 

At runtime, the Field Classifier is executed to identify the fundus images that are the most similar to the central and nasal fundus fields. Then, the Gradability and DR Classifiers are executed: While DR is evaluated in both the central and nasal fields, gradability is only evaluated in the central field, as it is the most representative for assessing DR gradability \cite{FieldsComparison2019, GradabilitySurvey2020}. If the eye screening request is formed by a single image, the Orchestrator assumes that it is a central fundus field, and evaluates both DR and gradability on that image.

The Orchestrator will return a referral proposal if either central or nasal DR output is positive\footnote{If both central and nasal fundus fields show DR signs, the Orchestrator will return the most severe DR scores.} (more than mild DR is detected) or if the central fundus field image is not gradable. The Orchestrator will return a non-referral proposal in all other cases. This behavior is consistent with international guidelines for DR screening, as a non-gradable eye fundus should always be referred \cite{Guidelines2017}. 

Note that the Orchestrator always assesses DR even if an image is classified as non-gradable. Thus, the output scores of DR and Gradability Classifiers (which are independent of each other), are always included in the screening proposal.

\paragraph{Field Classifier}
\label{sec:field-selector}
The Field Classifier is a neural network that classifies each fundus image into 7 custom categories, listed in Table \ref{fields-table} and previously illustrated in Figure \ref{fig:sample-studies}. We chose these categories because they adequately describe all images taken since the start of the screening program. Our categories do not require a distinction between the right and left eye (laterality), as the ETDRS imaging protocol does \cite{ETDRS1991}. In fact, the laterality of the eye is included as a tag in the DICOM file, so there is no need to use a model to identify the laterality of each fundus field.

NaIA-RD uses the Field Classifier assuming  that all the images of the request belong to the same eye. However, it does not impose any restriction on the imaging protocol: it works with any number of fundus images, ignoring non-standard ones. When multiple central and nasal fields are taken in an eye, this method usually selects the best quality image per category, as they resemble the most to the ideal fundus field.

\begin{table}
\begin{center}
\small
\setlength{\tabcolsep}{4pt}
\renewcommand{\arraystretch}{1.2}
\begin{tabular}{ lp{6.3cm}rr }
\toprule
Category & Description & Frequency & Samples \\ 
\midrule
Central & ETDRS field 2, centered on the macula. & 29.5\% & 28,634 \\ 
Nasal & ETDRS field 1, centered on the optical disk. & 29.5\%  & 28,749 \\ 
OD up & The image is centered on the inferior arcade, and the optical disk is visible. & 0.5\% & 489 \\ 
OD down & The image is centered on the superior arcade, and the optical disk is visible. & 0.5\% & 482 \\ 
No OD & Optical disk is not visible. & 3.9\% & 3,769 \\ 
Temporal & Temporal to the macula, while at least a portion of the optical disk is visible. & 1.7\% & 1,626  \\ 
Composite & Composition of multiple fields. & 34.4\%  & 33,464  \\ 
\bottomrule
\end{tabular}
\caption{\label{fields-table} Field selection model categories and their frequency (OD refers to Optical Disk).
}
\end{center}
\end{table}

\paragraph{Gradability Classifier}
\label{sec:grad-evaluator}
The Gradability Classifier is a neural network that classifies a central fundus field image as gradable/non-gradable, returning also a gradability score. This model produces a gradable (negative) output if at least 80\% of the eye fundus is visible \cite{GradabilitySurvey2020}. The gradability score is always included in NaIA-RD proposals as an image quality measurement.

\paragraph{DR Classifier}
\label{sec:dr-evaluator}
The DR Classifier is a neural network trained as a binary classifier to detect more than mild DR signs based on the ICDR scale. It assesses DR as referable/non-referable, classifying the fundus image as referable if it shows more abnormalities than just microaneurysms \cite{ICDR-2003}.

This model returns a negative output when the input image is deemed non-referable due to DR. This can occur if the fundus is barely visible in a non-gradable image. However, if a single hemorrhage is detected in a poor quality image, the DR Classifier will produce a positive output. For this reason, the DR score returned by this model is always included in NaIA-RD proposals as a measure of the DR signs present in the eye.

\subsubsection{Training techniques}
\label{sec:training-techniques}
All three neural networks mentioned above (Field Classifier, Gradability Classifier, and DR Classifier) are ResNet34 Convolutional Neural Networks \cite{ResNet2015}. We trained all of them using \texttt{fast.ai} v1 library \cite{Howard2020}, which provides a powerful wrapper of Pytorch. Each model uses square images of different sizes: 150px, 400px and 700px for Field, Gradability and DR Classifiers, respectively. We used a common preprocessing step for all models: finding and cropping the eye fundus circumference before the image is resized. We used standard \texttt{opencv} library functions for this task.

We initially loaded all the models with pretrained ImageNet weights, and progressively increased image size in each training iteration using progressive resizing \cite{Howard2020Book}. We also used \texttt{fast.ai}'s Batch Normalization, Weight Decay, MixUp and LabelSmoothing as regularization techniques. Additionally, we used several random image augmentations, such as: brightness and contrast adjusting, rotating, warping and cropping. 

We chose the best models based on their performance on the validation sets. The weights were saved at the end of each epoch when a maximum metric value was reached. For DR and Gradability Classifiers, we used the Area Under the Receiver Operating Characteristic Curve (AUROC, also known as AUC) metric, as a binary classifier metric that does not require a decision threshold. We obtained an AUC value of 0.979 for the DR Classifier, and an AUC value of 0.982 for the Gradability Classifier. For the Field Classifier, which is a multi-class model, we used the Cohen kappa metric rather than the AUC. This is due to kappa's superior simplicity for assessing multi-class classification problems \cite{ROCvskappa2008}. We obtained a Cohen kappa value of 0.976 in Field Classifier's validation set.

Then, we calibrated DR and Gradability Classifiers as we explain in Section \ref{sec:calibration}. Using the calibrated outputs, we chose the decision thresholds that maximized the arithmetic mean of the recalls of the positive and negative classes. We found a best threshold of 0.1 for the DR Classifier, which gave a sensitivity of 90.29\% and a specificity of 95.92\% on the validation set; and a threshold of 0.5 for the Gradability Classifier, which gave a sensitivity of 85.78\% and a specificity of 96.07\% on the validation set.

\subsubsection{Datasets}
\label{sec:data-sets}

We used 10 datasets to develop NaIA-RD, both public and private, which are summarized in Table \ref{data-sets-table}. The following subsections describe them, organized by the task they attempt to solve (field, DR, or gradability classification). The Gold Standard dataset, which we consider to be the clinical reference standard for NaIA-RD, is also presented.

\begin{table}
\small
\begin{center}
\begin{threeparttable}
\setlength{\tabcolsep}{4pt}
\begin{tabular}{ @{}llp{3cm}c*3r@{} }
\toprule
Dataset & Source & Components & Classes & \multicolumn{3}{c}{Image quantity} \\
{} & {} & {} & {} & \textit{Train} & \textit{Valid} & \textit{Test} \\
\midrule
Fundus Field & Private & Field Classifier & 7 & 36,317 & 29,968 & 29,928 \\
Gradability & Private & Gradability Classifier & 2 & 8,744 & 8,592 & -- \\
DR & Private & DR Classifier & 2 & 17,336 & 8,592 & -- \\
Gold Standard & Private & NaIA-RD \newline DR Classifier \newline Gradability Classifier & 3 & -- & -- & 984 \\
EyePACS (Kaggle) \cite{KaggleEyePACSCompetition2016} & Public & DR Classifier & 2 & 77,787 & -- & 10,906 \\
APTOS (Kaggle) \cite{APTOSKaggle2019} & Public & DR Classifier & 2 & 1,831 & -- & 1,831 \\
Messidor-2 \cite{Messidor2018,Messidor-2Data} & Public & DR Classifier & 2 & 872 & -- & 872 \\
IDRiD \cite{IDRIDDataset} & Public & DR Classifier & 2 & 413 & -- & 103 \\
OIA-DDR \cite{OIA-DDR-2019} & Public & DR Classifier \newline Gradability Classifier & 3 & -- & -- & 4,105 \\
EyeQ \cite{EyeQFu2019} & Public & Gradability Classifier & 2 & -- & -- & 16,249 \\
\bottomrule
\end{tabular}
\end{threeparttable}
\caption{\label{data-sets-table} Summary of datasets used for NaIA-RD development. Note that the Gold Standard was used to test multiple components. Also, some datasets are not used for training, validation, or testing. For example, OIA-DDR and EyeQ datasets are used for testing only.
}
\end{center}
\end{table}

\paragraph{Field classification}
We used a private dataset to train, validate, and test the Fundus Field Classifier neural network. We classified more than 96,000 HUN images into one of the 7 fundus fields listed in Table \ref{fields-table}. We detail the obtained test metrics in Section \ref{sec:field-test-set}.

\paragraph{DR classification}
\label{sec:dr-classification-data-set}
We used the following public and private datasets to train, validate and test the DR classification model:

\begin{enumerate}
    \item \textit{Private datasets}: We created a private dataset (DR dataset), labeled by our engineering team, to train and select the best DR Classifier model (validation). We ensured that its 25,928 images were labeled as referable/non-referable based on visible DR signs, strictly following the ICDR grading standard \cite{ICDR-2003}. In addition, the DR Classifier was carefully tested using the Gold Standard dataset (Section \ref{sec:gold-standard}).
    \item \textit{Public datasets}: We used 4 popular public datasets to train and test the DR Classifier: EyePACS from the Kaggle 2016 competition \cite{EyePACS2009, KaggleEyePACSCompetition2016, KaggleEyePACSFulldatasets}, APTOS from the Kaggle 2019 competition \cite{APTOSKaggle2019}, Messidor-2 \cite{Messidor2018,Messidor-2Data} and IDRiD \cite{IDRIDDataset}. Additionaly, we used the OIA-DDR dataset \cite{OIA-DDR-2019} for external validation, meaning that its images were used only to test the generalizability of the model.
\end{enumerate}

\begin{figure}
\small
\centering
\includegraphics[width=0.5\linewidth]{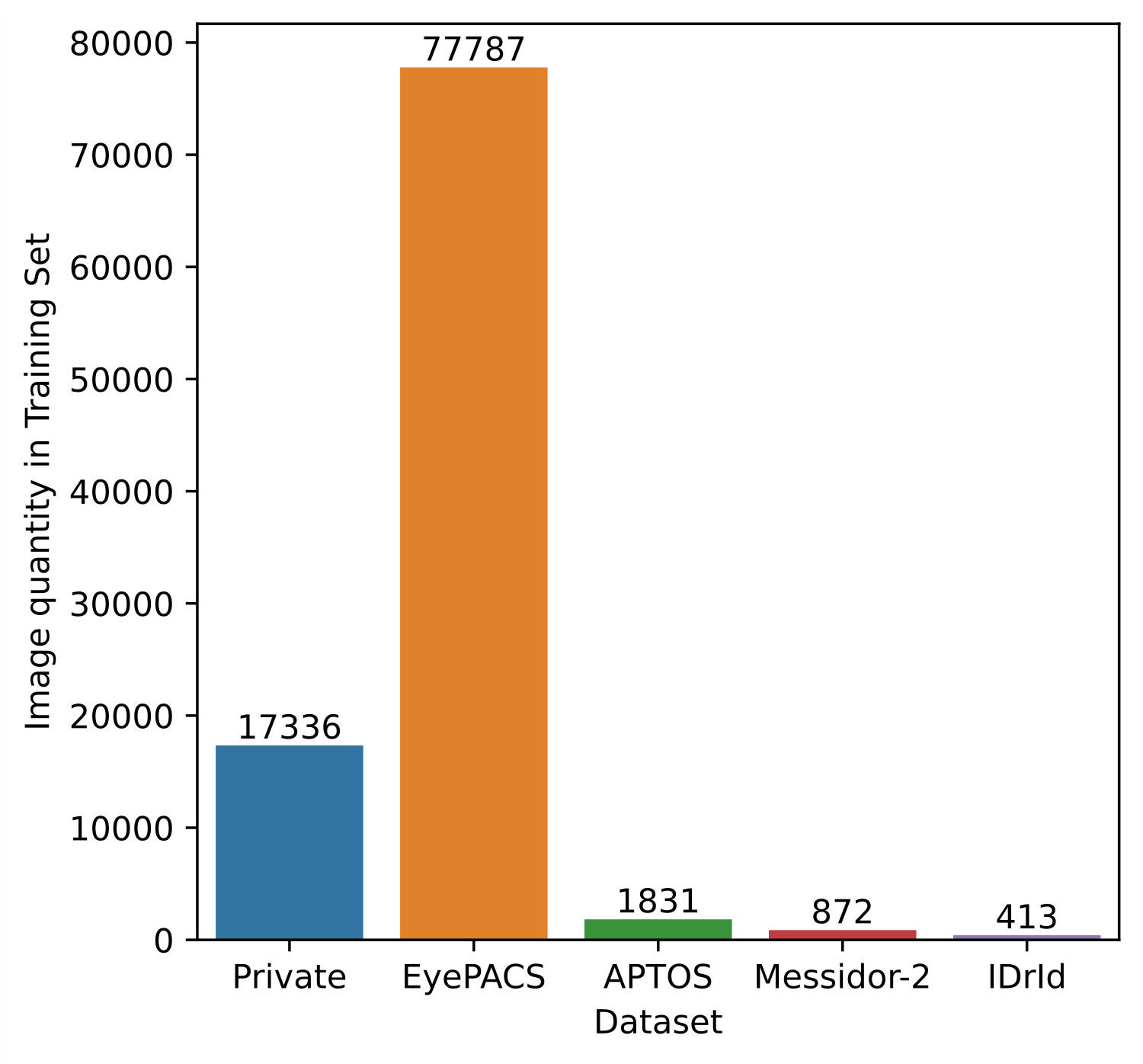}
\caption{Source of the data used to train the DR Classifier.}
\label{fig:data-sets-in-train}
\end{figure}

We trained the DR Classifier jointly using the aforementioned public and private datasets, excluding the Gold Standard and OIA-DDR. Figure \ref{fig:data-sets-in-train} shows that most of this training set comes from EyePACS, while private data is only 17.65\% of the total. These datasets involve diverse patient populations, cameras and imaging conditions. Using multiple datasets for training DR classifiers is a common approach \cite{DeepLearningDRSurvey2022}. However, we chose the DR Classifier with the best AUC in our private validation subset (0.979), as it represented the target data distribution.

We also tested the DR Classifier using the same public datasets we used for training. Specifically, we used the public test sets of EyePACS Kaggle 2016 and IDRiD. As we could not find any defined test set, we used a random 50\% sample of APTOS Kaggle 2019 and Messidor-2 datasets. Since we found some incorrect labels in the APTOS dataset, we relabeled some \textit{mild DR} images (9\%) as \textit{moderate DR} (referable), according to the ICDR severity scale \cite{ICDR-2003}.

Finally, we evaluated DR Classifier's generalization capabilities using the OIA-DDR dataset, which was totally excluded from the training process. OIA-DDR is formed by 13,000 labeled fundus images of 9,500 patients, which were taken in 147 hospitals using 42 different fundus camera models (Topcon D7000, Topcon NW48, Nikon D5200, Canon CR 2 and others). All images were labeled by four professional graders. The authors propose a random 30\% of this data as a test set, from which we excluded 346 images labeled as non-gradable. Our final test set consisted of 3,759 images, where 1,690 were labeled as referable (45\%). We used the non-gradable images to assess the Gradability Classifier, as we explain in Section \ref{sec:gradability-datasets}.

All these public datasets follow the ICDR severity scale for DR grading, with DR grades ranging from 0 to 4 (\textit{0 - No DR, 1 - Mild, 2 - Moderate, 3 - Severe, 4 - Proliferative DR}). We binarized them as referable/non-referable DR, considering more than mild grades (grade > 1) as referable. 

We detail the obtained metrics in Section \ref{sec:external-validation}, while in Section \ref{sec:discussion} we compare them with the results published in other works.

\paragraph{Gradability classification}
\label{sec:gradability-datasets}
We used the following public and private datasets to train, validate and test the gradability classification model:

\begin{enumerate}
    \item \textit{Private datasets}: We created a dataset consisting of more than 17,000 images from HUN (gradability dataset) to train and validate the Gradability Classifier. We labeled each image as gradable or non-gradable under close expert supervision.
    \item \textit{Public datasets}: To test the Gradability Classifier with external data, we used two public datasets that were totally excluded from training and validation: OIA-DDR \cite{OIA-DDR-2019} and EyeQ \cite{EyeQFu2019}.
\end{enumerate}

We trained and chose the best performing Gradability Classifier model using our private dataset. Our grading criteria were based on fundus visibility: a gradable fundus image should allow localization of small hemorrhages, especially in the macular area. We considered image focus, clarity, artifacts, macular visibility, and the gradable area, which had to reach 80\% of the image \cite{GradabilitySurvey2020}. Our labellers were allowed to make brightness and contrast adjustments in order to ignore easily fixable image quality issues. Note that both the gradability and DR classifiers share the same validation images, but the gradability training set is a subset of the DR training set. We respected these partitions to avoid any system-level bias.

Later, we evaluated the generalization capabilities of the Gradability Classifier using the OIA-DDR test set \cite{OIA-DDR-2019}. OIA-DDR is a large public dataset introduced in Section \ref{sec:dr-classification-data-set}. This dataset consists of 346 non-gradable and 3,759 gradable images in which a DR severity grade has been assigned. Therefore, we used the non-gradable category to obtain a binarized dataset suitable for external validation.

Nevertheless, OIA-DDR is not specifically designed to assess retinal image quality. For this purpose, we used the Eye Quality Assessment Dataset (EyeQ) \cite{EyeQFu2019}. In EyeQ, two experts labeled 28,792 retinal images from the EyePACS (Kaggle) \cite{EyePACS2009} image set in three categories: \textit{good}, \textit{usable}, and \textit{reject}. The \textit{reject} category indicates that the image is not suitable for a reliable diagnosis. Therefore, we binarized the EyeQ test set based on this category, resulting in 3,215 non-gradable and 13,029 gradable images. 

\paragraph{Gold Standard}
\label{sec:gold-standard}
The Gold Standard is a private dataset we created to evaluate NaIA-RD as a black box prior to its deployment. It is labeled by expert ophthalmologists from HUN. We used it as a clinical reference standard to evaluate the overall DR screening capabilities of the system, as well as two of its key components: DR and Gradability classifiers. We report the obtained metrics in Section \ref{sec:gold-standard-res}.

To create this dataset, we first agreed with HUN to measure an expected sensitivity of 80\% with a confidence interval (CI) width of 10\% and a confidence level of 95\%. Given an estimated prevalence of referable eyes of 7\%, we calculated a required dataset size of 1,265 eyes to measure the expected sensitivity \cite{SampleSizeBurderer1996, SampleSizeHajianTilaki2014}. However, we decided to reduce the required labeling effort by artificially increasing the prevalence (but maintaining the same statistical properties), resulting in a dataset size of 492 eyes with a prevalence of 18\% referable eyes.

Therefore, we selected a sample of 492 eyes (retinographies, which we will refer to as \textit{eyes} for simplicity) using anonymized clinical records (from March 2019 to August 2019), ensuring a prevalence of 18\% of referable eyes. These eyes belonged to 205 different patients\footnote{When we randomly selected an eye for the Gold Standard, we added the corresponding fellow eye (if present) to complete an entire study. We will use the term study to refer to an object formed by fundus images of two fellow eyes. Following this procedure, some patients were represented in the Gold Standard with multiple studies due to the random nature of eye selection, but we ensured that the same study was not included twice.}, 66\% males and 34\% females (sex assigned at birth), with a mean age of 64.25 years (SD 14.87) at the time of the study. We excluded all these patients from all training and validation sets of NaIA-RD.

Three ophthalmologists from the hospital labeled each eye. Each expert provided a blind, independent label (non-referable, referable due to non-gradable fundus or referable due to DR), following the ICO guidelines and the ICDR scale as grading standard \cite{Guidelines2017, ICDR-2003}. For each eye, the expert visualized all the fundus images taken by the nurse during the imaging session, along with an enhanced version created using the image enhancement technique detailed in Section \ref{sec:image-enhancement}. After labeling was completed, we discarded eyes that had received three different votes (no consensus). 3 eyes were discarded by this procedure, resulting in a final Gold Standard dataset of 489 eyes.

We deployed NaIA-RD in a simulated production environment, obtaining its output for each eye of the dataset. Each eye was composed of the same real-world fundus images that the experts had labeled. Using this procedure, we compared the returned screening proposals with the expert labels and evaluated NaIA-RD in three different tasks: DR screening, DR classification, and gradability classification.

\vspace{6pt}
\noindent\textbf{Task 1: DR Screening.} We evaluated NaIA-RD for binary DR screening (refer/not refer), without taking the motivation into account (DR or non-gradability). We compared the performance of NaIA-RD in three different ways using this data: 
\begin{enumerate}
    \item \textbf{Compared with the consensus of 3 ophthalmologists}. The main goal of the Gold Standard was to compare NaIA-RD with the best possible clinical judgement, so we compared NaIA-RD's proposal per eye with the simple majority label of the experts (refer/not refer). As 3 ophthalmologists had graded all the eyes, no ties were possible.
    \item \textbf{Compared with a single ophthalmologist}. We also wanted to compare NaIA-RD with an ophthalmologist performing the screening alone. To do this, we obtained the metrics of each Gold Standard labeler using the majority label of the other two ophthalmologists as the ground truth. We used NaIA-RD's outputs to break ties. Then, we compared NaIA-RD's metrics with those of each ophthalmologist.
    \item \textbf{Compared with first-level screening GPs in real-world settings}. We compared both NaIA-RD proposals and GP decisions in the screening program with the majority label of the ophthalmologists. DR screening decisions had been made per patient, involving both eyes, so for this comparison we used the positive class (refer) if one of the eyes was considered referable (both the Gold Standard and NaIA-RD provided a label per eye). Due to missing data in clinical records, we had to use a subset of the Gold Standard for this comparison. The final dataset consisted of 122 screening decisions involving 244 eyes.
\end{enumerate}

\vspace{6pt}
\noindent\textbf{Task 2: DR classification.} We evaluated NaIA-RD for referable/non-referable DR classification without considering gradability. Our goal was to test the DR Classifier model isolated from the Gradability Classifier model. Therefore, we first discarded eyes graded as non-gradable by simple majority (15 eyes), and we finally used the resulting majority label as the ground truth. We did not find any ties using this procedure. We binarized NaIA-RD outputs considering a positive class only if NaIA-RD outputted referable due to DR.

\vspace{6pt}
\noindent\textbf{Task 3: Gradability classification}. Analogously, we evaluated NaIA-RD for gradable/non-gradable classification without considering DR. Eyes with a non-gradable simple majority vote were taken as non-gradable (15 eyes), otherwise they were taken as gradable. We binarized NaIA-RD outputs considering a positive class only if NaIA-RD outputted referable due to non-gradability.

\subsubsection{Calibration}
\label{sec:calibration}
Models whose scores are to be used for human decision making or automation should be calibrated \cite{Calibration2021}. This is the case for NaIA-RD, where the output scores will be interpreted by clinicians and the HIS. Therefore, we calibrated both the DR and Gradability classifiers, while the Field Classifier did not need this feature. 
We have addressed model calibration as the next step after selecting the best model (this is called \textit{post-hoc} calibration \cite{Calibration2021}). 



To train two separate DR and Gradability calibrators, we first tried using their respective training sets, but they performed poorly. So, we used their entire validation sets for this purpose. We chose the Beta Calibration \cite{BetaCalibrationOriginal-Kull-2017} technique for the DR classifier, and Isotonic Regression \cite{IsotonicOriginal2002-Zadrozny} for the Gradability Classifier, based on systematic cross-validation experiments on the validation sets. The estimated calibration error, mean calibration error and Brier Score of the uncalibrated models were 0.017248, 0.299778, 0.023504 (DR) and 0.114798, 0.278866, 0.063906 (Gradability), respectively (mean values), and the calibrated models obtained mean values of 0.005958, 0.155110, 0.021976 (DR) and 0.010228, 0.116718, 0.045477 (Gradability), respectively. More details along with calibration curves are included in the Supplementary Material.

Unfortunately, presenting two separate calibrated probabilities in a screening proposal may be difficult to interpret. It would be more convenient to combine the DR and gradability scores into a single number. Additionally, the DR and gradability scores should be self-explanatory, meaning it should not be necessary to know the decision threshold to understand them. For example, it would be counterintuitive to suggest a referral for possible DR with only a 10\% probability (where the DR referral threshold is set to 0.1).. 

Therefore, the scores returned by NaIA-RD are not the calibrated probabilities directly. Instead, given a new decision boundary $t'$, NaIA-RD returns a transformed referral score $s$, where $s < t'$ is given for non-referral proposals, and $s \geq t'$ for referrals. We used $t'=0.5$ in order to make the screening score more intuitive. Given a probability $p$, a threshold $t$ and a decision boundary $t'$, we have defined a transformation function $f\left(p, t, t'\right)$ as follows: 

\[
    f\left(p, t, t'\right)= 
\begin{cases}
    t'\cdot{\left(1+\frac{p - t}{1 - t}\right)},& \text{if } p > t\\
    t'\cdot{\left(1+\frac{p - t}{t}\right)},& \text{otherwise}
\end{cases}
\]
This transformation has the property of preserving the order of the input probabilities $p$, applying a different linear transformations for each case. After transforming the DR and gradability scores, NaIA-RD generates a single DR screening score, using the highest transformed value from both classifiers in its proposal.

\subsubsection{Interpretability}

The DR classifier generates heatmaps for positive predictions to improve interpretability. We understand model interpretability as the intuitive mapping between inputs and outputs, as described in \cite{ExplainableAI2020Linardatos}. Therefore, we used a technique called Integrated Gradients \cite{IntegratedGradientsMethod-Sundararajan-2017} to be able to highlight small lesions on the input image. This technique provides an accurate mapping of input image pixels without modifying the original neural network by approximating the integral of the gradients of neural activations. Integrated Gradients was successfully used for DR screening interpretability in \cite{IntegratedGradientsExplainability-Sayres-2019}. The authors concluded that heatmaps can increase the confidence and accuracy of human graders, but also their grading time, so heatmaps should only be used for positive DR predictions.

NaIA-RD uses this attribution technique to provide interpretability: First, the Integrated Gradients algorithm is applied to obtain an attribution mask (Gauss-Legendre integral approximation method is executed for 20 forward steps using the \texttt{captum} library \footnote{Captum library: \url{https://captum.ai/api/integrated_gradients.html}}. Then, the attribution mask pixels are clustered using the OPTICS algorithm \cite{Optics1999} (\texttt{sklearn}'s \texttt{OPTICS} class is used, with an \textit{epsilon} maximum distance between two cluster points of 23 and a \textit{min\_samples} minimum cluster size of 4). Finally, NaIA-RD returns the center coordinates and radios of the clusters, which the HIS overlays on the image as standard DICOM circumference annotations. These circumferences usually highlight DR signs such as hemorrhages. In Figure \ref{fig:ret-expl} we show an example that illustrates this process.

\begin{subfigure*}
\setcounter{subfigure}{0}
    \centering
    \begin{minipage}[b]{0.3\textwidth}
        \includegraphics[width=.99\textwidth]{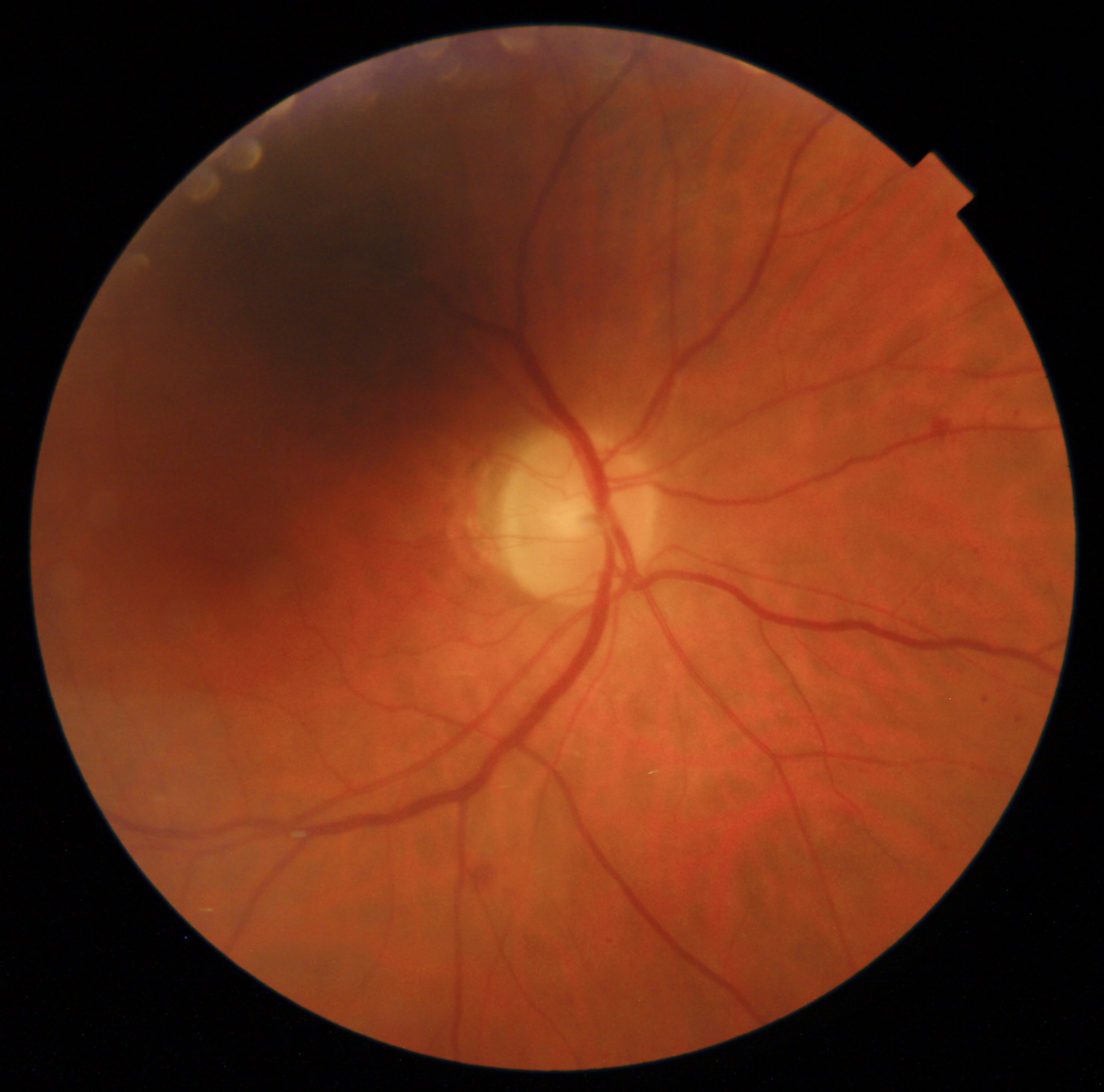}
        \caption{Original image.}
    \end{minipage}
    \begin{minipage}[b]{0.3\textwidth}
        \includegraphics[width=.99\textwidth]{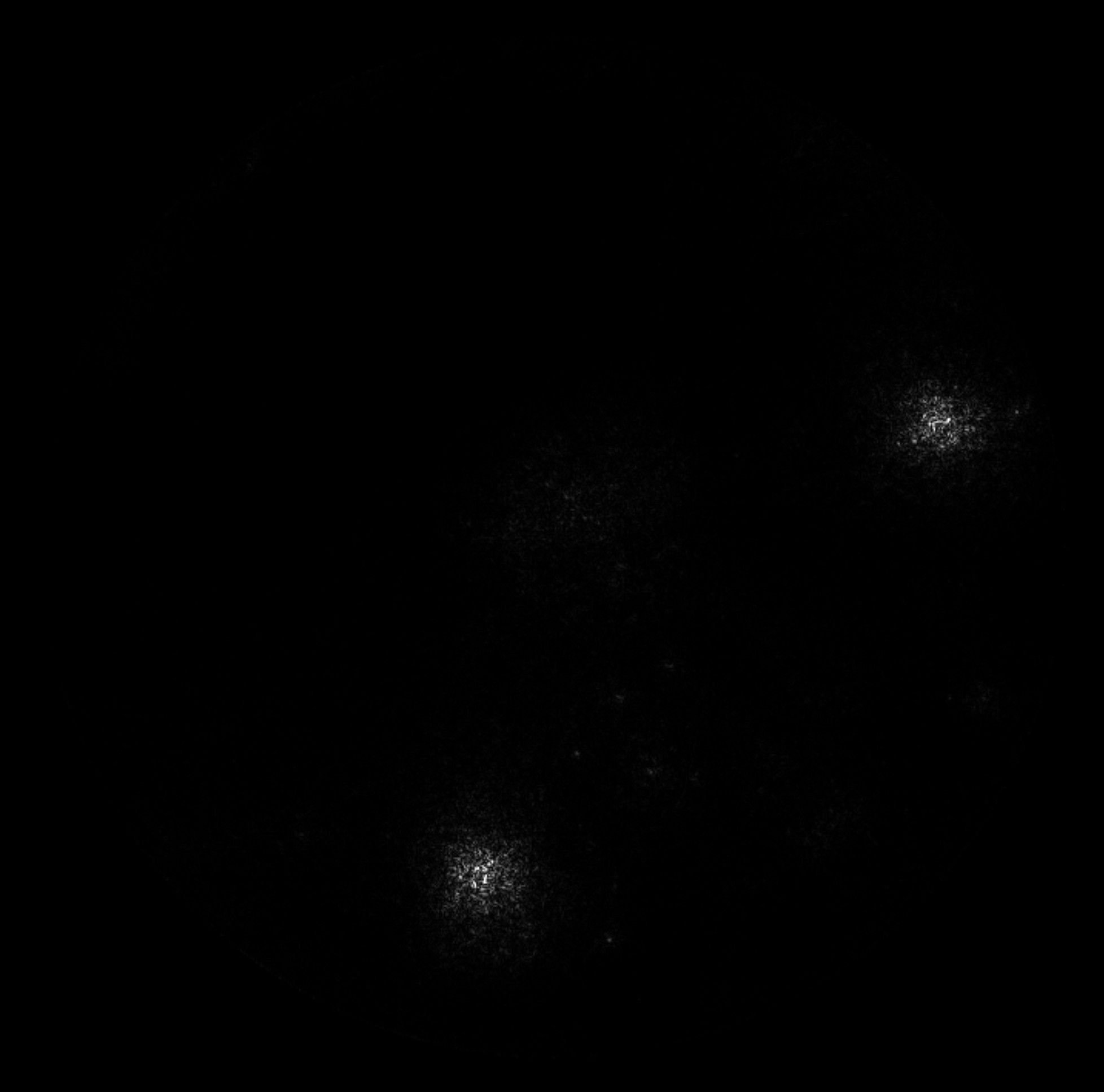}
        \caption{Pixel attribution.}
    \end{minipage}
    \begin{minipage}[b]{0.3\textwidth}
        \includegraphics[width=.99\textwidth]{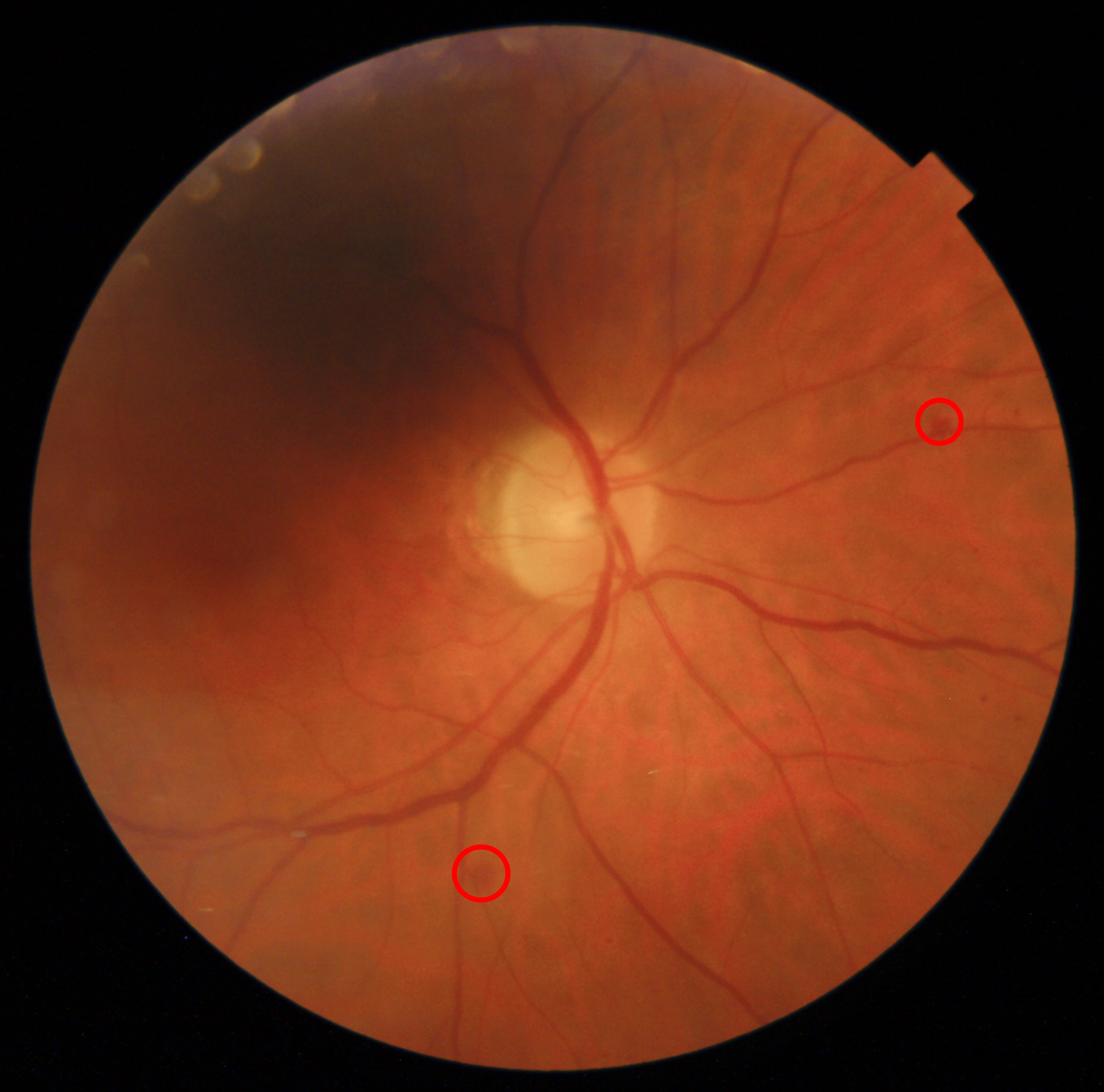}
        \caption{Annotations.}
    \end{minipage}

    \setcounter{subfigure}{-1}
    \caption{Annotation process of NaIA-RD using Integrated Gradients, as a mechanism of increasing interpretability. NaIA-RD provides pixel attributions as circumference coordinates, which the HIS stores in the DICOM object as standard DICOM annotations.}
    \label{fig:ret-expl}
\end{subfigure*}

\subsubsection{Image enhancement}
\label{sec:image-enhancement}
In addition to the DR screening proposal, NaIA-RD provides an enhanced image of the eye fundus to ease human interpretation. In particular, it returns two enhanced central and nasal images per request, and the HIS stores them in the DICOM study. In this way, clinicians can benefit from image enhancement using any DICOM viewer.

In Figure \ref{fig:vnoct} we show an example of how NaIA-RD enhances a challenging fundus image. The enhanced image is computed in three steps: 
\begin{enumerate}
    \item The eye fundus is cropped from the original image.
    \item The dynamic range of the image is linearly extended after clipping intensity values below 1\% and above 99\% percentiles, respectively.
    \item Contrast is further increased applying Contrast Limited Adaptive Histogram Equalization (CLAHE) \cite{Clahe1987, Clahe1994} in each RGB channel. We use \texttt{opencv}'s \texttt{createCLAHE} function for this last computation. 
\end{enumerate}

We have found that extending the dynamic range (step 2) before applying CLAHE (step 3) gives better results than applying CLAHE first.


\setcounter{figure}{6}
\begin{subfigure*}
    \setcounter{subfigure}{0}
    \centering
    \begin{minipage}[b]{.33\textwidth}
        \includegraphics[width=.99\textwidth]{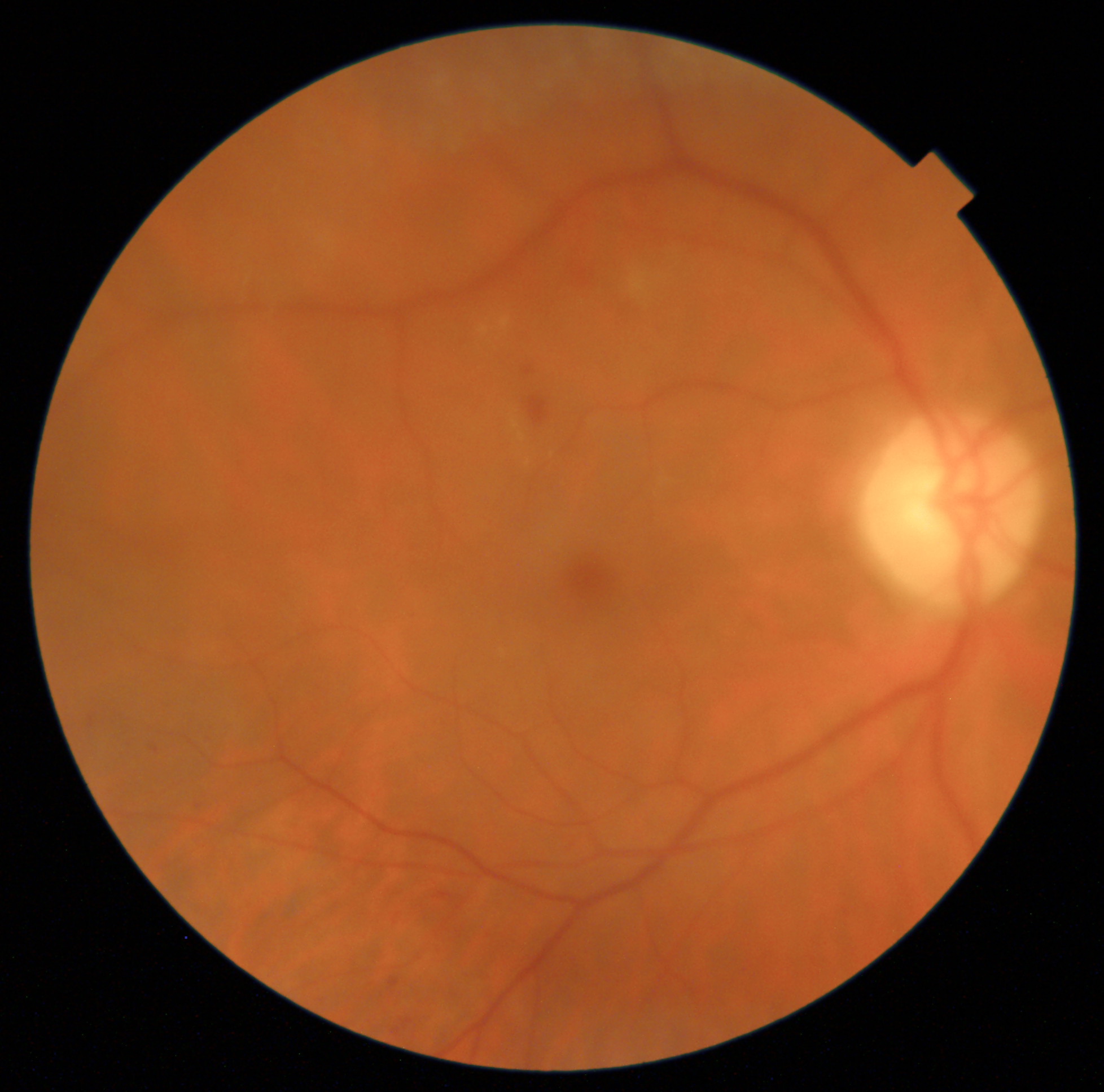}
        \caption{Original.}
    \end{minipage}
    \begin{minipage}[b]{.33\textwidth}
        \includegraphics[width=.99\textwidth]{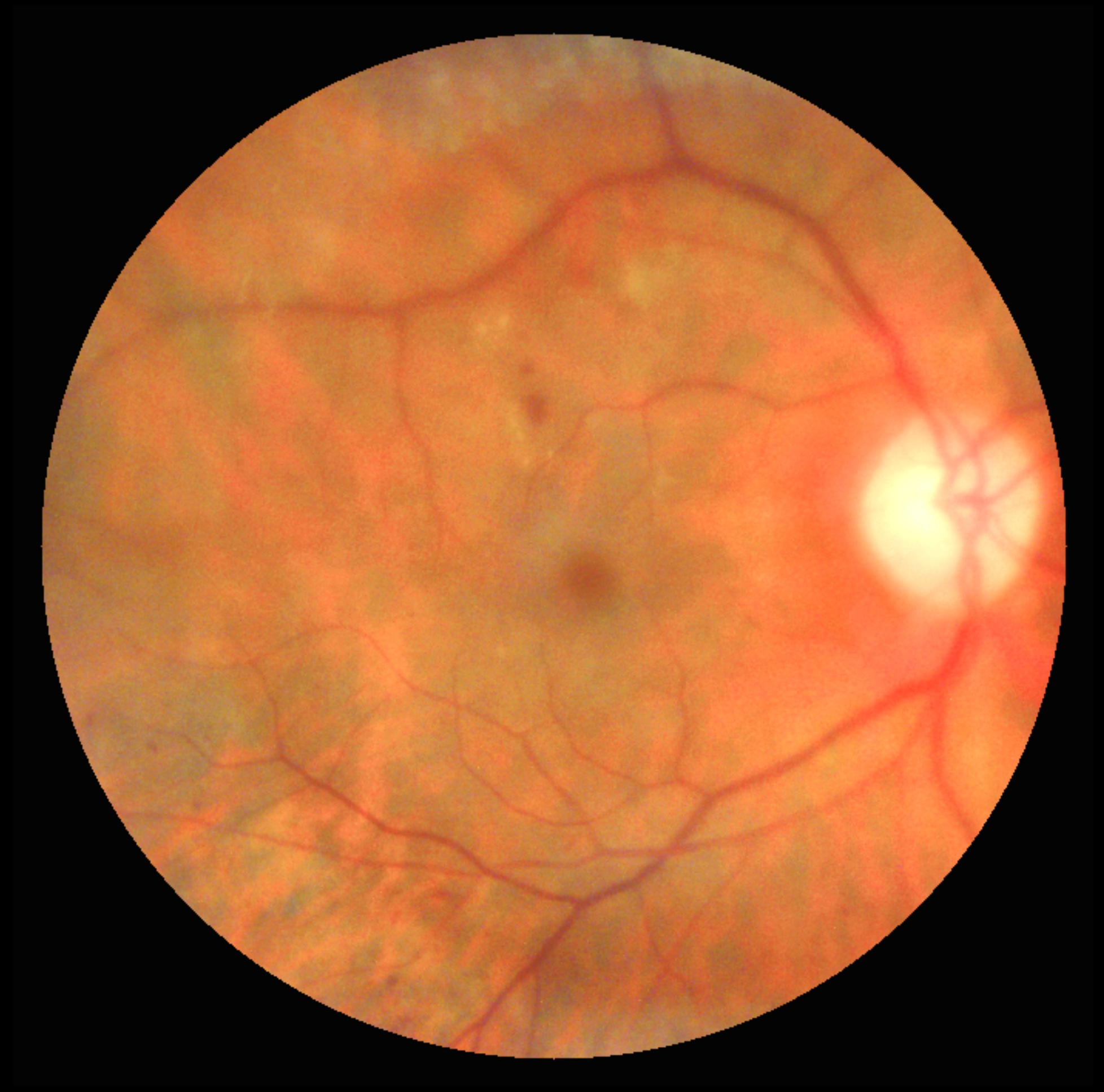}
        \caption{Enhanced.}
    \end{minipage}
    \setcounter{subfigure}{-1}
    \caption{Example of fundus image enhancement by NaIA-RD. Notice how two hemorrhages near the macula are more visible in the enhanced image. Another smaller bleedings are also more evident in the inferior arcade.}
    \label{fig:vnoct}
\end{subfigure*}

\subsubsection{MLOps}
\label{sec:mlops}
NaIA-RD was developed in Python, with each component providing a REST API and running in its own Docker container (see Figure \ref{fig:components}). All code was written in Jupyter notebooks, using the \texttt{nbdev} \footnote{\texttt{nbdev} library: \url{https://nbdev.fast.ai/}} library to implement a literate programming paradigm \cite{LiterateProgramming1992-Knuth}. We extensively unit tested all code and components, and a sanity check job was run periodically to notify if any significant deviation was detected in the last month's data.

\section{Results}
\label{sec:results}

In this section we evaluate NaIA-RD using the datasets we introduced in Section \ref{sec:data-sets}. The performance of NaIA-RD is compared with that of experts, and its ability to generalize is assessed. These results were important to HUN, as the decision to deploy NaIA-RD was based on them.

Additionally, this section presents the before-and-after study we conducted. This study compares the screening decisions made at HUN before (retrospectively) and after (prospectively) the deployment of NaIA-RD, measuring its real-world impact on clinicians and patients.

When appropriate, we report the area under the ROC curve (AUC), the sensitivity and the specificity, as similar works do  \cite{Abramoff2018USProspectiveIDX, IDXPivotalTrial2018, Sha2020ValidationIdxValencia, Lim2022ProspectiveMulticenterEyeArt, Ribeiro2014ProspectiveRetmarker, OpthAIArxivQuellec2019}. We also use the Cohen kappa score to measure agreement \cite{KappaMcHugh2012}. All confidence intervals are calculated with a 95\% of confidence level using bootstrapping \cite{mooney1993bootstrapping}.

\subsection{Field classification}
\label{sec:field-test-set}
The Field Classifier obtains a weighted multi-class AUC of 99.62\% on the test set (20,858 fundus images that were not used for optimization nor model selection). 

\subsection{Gold Standard}
\label{sec:gold-standard-res}
As described in Section \ref{sec:gold-standard}, we used our private Gold Standard to evaluate NaIA-RD on three different tasks: DR screening, DR classification, and gradability classification. Table \ref{gs-naiard-vs-consensus} shows the overall results of NaIA-RD on these tasks, while Tables \ref{gs_ophthalmologists} and \ref{gs_GPs_vs_naiard_table} provide a deeper comparison between NaIA-RD and individual ophthalmologists as well as first-level screening GPs. The most relevant results for each task are summarized below.

\vspace{6pt}
\noindent\textbf{Task 1: DR screening}.
\begin{enumerate}
    \item \textbf{Compared with the consensus of 3 ophthalmologists}. According to the first row of Table \ref{gs-naiard-vs-consensus}, NaIA-RD achieves a sensitivity and specificity greater than 92\%, with a Cohen kappa of 0.81 (strong agreement \cite{KappaMcHugh2012}).
    \item \textbf{Compared with a single ophthalmologist}. In Table \ref{gs_ophthalmologists} we can observe that NaIA-RD is the only grader with sensitivity and specificity above 91\%, and its Cohen kappa score (0.818) is higher than the kappa score of the two thirds of the ophthalmologists.
    \item \textbf{Compared with first-level screening GPs in real-world settings}. Table \ref{gs_GPs_vs_naiard_table} shows a significantly lower sensitivity with a much wider confidence interval for GPs (26.6\%-63.3\%) than for NaIA-RD (83.3\%-100\%). Cohen kappa scores show a weak agreement (0.432) for GPs, while NaIA-RD shows a moderate-strong agreement (0.794).
\end{enumerate}

\vspace{6pt}
\noindent\textbf{Task 2: DR classification task}. The second row of Table \ref{gs-naiard-vs-consensus} shows that the performance of NaIA-RD on this task is superior to its performance on Task 1: DR screening (the obtained AUCs are 0.986 and 0.979, respectively).

\vspace{6pt}
\noindent\textbf{Task 3: Gradability classification}. The third row of Table \ref{gs-naiard-vs-consensus} shows a gradability classification specificity of 97.2\% (95.5-98.5). However, only 15 eyes (3\%) were graded by the ophthalmologists as non-gradable: this resulted in a wide confidence interval for the sensitivity measurement (26.6-80.0), revealing a limitation of this dataset.

\begin{table*}[htbp]
\renewcommand{\arraystretch}{1.5}
\centering
\small
\resizebox{0.98\textwidth}{!}{
\begin{tabular}{ @{}clp{3cm}p{1.65cm}p{1.6cm}p{1.6cm}p{1.6cm}p{1.6cm}p{1.6cm} @{}}
\toprule
ID & Task & Target question & Ground truth positives & Kappa & AUC & Sensitivity & Specificity  \\ 
\midrule
1 & DR screening & Is referable to the ophthalmologist? & 27.61\% \newline 135/489 & 0.818 \newline 0.764-0.869 & 0.979 \newline 0.969-0.989 & 92.5\% \newline 88.1-96.3 & 92.4\% \newline 89.5-94.9 \\
2 & DR classification & Are there observable signs of more than mild DR? & 25.3\% \newline 120/474 & 0.852 \newline 0.801-0.901 & 0.986 \newline 0.977-0.993 & 91.6\% \newline 86.6-95.8 & 95.2\% \newline 92.9-97.2 \\
3 & Gradability classification & Is non-gradable for DR screening? & 3\% \newline 15/489 & 0.423 \newline 0.229-0.602 & 0.754 \newline 0.684-0.822 & 53.0\% \newline 26.6-80.0 & 97.2\% \newline 95.5-98.5 \\
\bottomrule
\end{tabular}}
\caption{\label{gs-naiard-vs-consensus} NaIA-RD metrics on different tasks using our private Gold Standard. Ground truth labels are obtained as the majority label of three ophthalmologists. For the DR classification task, non-gradable images are excluded. All confidence intervals are calculated with a 95\% of confidence level using bootstrapping \cite{mooney1993bootstrapping}.
}
\end{table*}

\begin{table*}
\renewcommand{\arraystretch}{1.5}
\centering
\small
\begin{tabular}{@{} rp{1.65cm}p{1.8cm}p{1.8cm}p{1.8cm} @{}}
        \toprule
         & Ground truth positives & Kappa & Sensitivity & Specificity  \\ 
        \midrule
        Ophthalmologist 1 & 28.66\% \newline 141/489 & 0.835 \newline 0.781-0.89 & 89.4\% \newline 84.2-94.3 & 94.7\% \newline 92.2-96.8 \\
        Ophthalmologist 2 & 30.47\% \newline 149/489 & 0.805 \newline 0.753-0.861 & 78.4\% \newline 71.4-84.6 & 98.2\% \newline 96.8-99.4 \\
        Ophthalmologist 3 & 27.81\% \newline 136/489  & 0.789 \newline 0.736-0.844 & 93.3\% \newline 88.9-97.1 & 90.0\% \newline 86.9-92.3 \\
        NaIA-RD & 27.61\% \newline 135/489 & 0.818 \newline 0.764-0.869 & 92.5\% \newline 88.1-96.3 & 92.4\% \newline 89.5-94.9 \\
        \bottomrule
    \end{tabular}
    \caption{\label{gs_ophthalmologists} DR screening performance comparison per ophthalmologist versus the majority label of the other two ophthalmologists of the Gold Standard and NaIA-RD. All confidence intervals are calculated with a 95\% of confidence level using bootstrapping \cite{mooney1993bootstrapping}.}
\end{table*}

\begin{table*}
\renewcommand{\arraystretch}{1.5}
\centering
\small
\begin{tabular}{ @{}rp{1.65cm}p{1.8cm}p{1.8cm}p{1.8cm} @{}}
        \toprule
        & Ground truth positives & Kappa & Sensitivity & Specificity  \\ 
        \midrule
        GPs & 24.59\% \newline 30/122 & 0.432 \newline 0.241-0.626 & 43.2\% \newline 26.6-63.3 & 94.4\% \newline 89.1-98.9 \\
        NaIA-RD & 24.59\% \newline 30/122 & 0.794 \newline 0.674-0.913 & 93.6\% \newline 83.3-100.0 & 91.3\% \newline 85.3-96.7\\
        \bottomrule
    \end{tabular}
    \caption{\label{gs_GPs_vs_naiard_table} First-level screening GPs (GPs) and NaIA-RD metrics on a subset of the Gold Standard. Real-world screening decisions from four different GPs are used as labels. Each screening decision involves two eyes (right and left eye). All confidence intervals are calculated with a 95\% of confidence level using bootstrapping \cite{mooney1993bootstrapping}.}
\end{table*}

\subsection{External validation}
\label{sec:external-validation}
\noindent\textbf{DR classification}. Table \ref{external-metrics-table} shows the results of the DR Classifier over the test partitions of several public datasets. The obtained AUCs range from 0.957 to 0.999. Note that the public test sets of EyePACS, IDRiD, and OIA-DDR can be used directly for comparison with previous works. In the discussion section (Section \ref{sec:performance}), we make this comparison. However, recall from Section \ref{sec:dr-classification-data-set} that some EyePACS and IDRiD partitions were used for training, so these metrics should be used with caution.

On the contrary, OIA-DDR was totally excluded from training. In this dataset, NaIA-RD obtained an AUC of 0.957, a Cohen kappa of 0.76 (moderate agreement), and a sensitivity and specificity of 93\% and 84\%, respectively. 

\vspace{6pt}
\noindent\textbf{Gradability classification}. On the OIA-DDR test set, the Gradability Classifier obtained an AUC of 0.928, a Cohen kappa of 0.213 (minimal agreement), and sensitivity/specificity of 100\% and 62\%, respectively. When assessing the EyeQ test set, the Gradability Classifier obtained an AUC of 0.938, Cohen kappa of 0.63 (moderate agreement), and sensitivity/specificity of 87\% and 86\%, respectively.

\begin{table}
\centering
  \small
  \renewcommand{\tabcolsep}{4pt}
 \begin{tabular}{rlp{6cm}rcrr}
        \toprule
         Dataset & Reference & Test partition & Size & Relabeled & AUC & Kappa \\ 
        \midrule
        EyePACS (Kaggle) & \cite{EyePACS2009, KaggleEyePACSCompetition2016, KaggleEyePACSFulldatasets} & Public test set & 10,906 & No & 0.964 & 0.818 \\
        APTOS (Kaggle) & \cite{APTOSKaggle2019} & 50\% random & 1,831 & Yes\tnote{1} & 0.999 & 0.962 \\
        Messidor-2 & \cite{Messidor2018,Messidor-2Data} & 50\% random & 872 & No & 0.988 & 0.877 \\
        IDRiD & \cite{IDRIDDataset} & Public test set & 103 & No & 0.991 & 0.879 \\
        OIA-DDR & \cite{OIA-DDR-2019} & Public test set without non-gradable images & 3,759 & No & 0.957 & 0.760 \\
        \bottomrule
    \end{tabular}
    \begin{tablenotes}
        \item[1] We relabeled 165 images (9\%) from \textit{mild} to \textit{moderate DR} using the ICDR \cite{ICDR-2003} scale.
    \end{tablenotes}
    \caption{\label{external-metrics-table} NaIA-RD metrics on external datasets assessing DR classification.}
\end{table}

\subsection{Before-and-after study}
\label{sec:prospective-study}
Four trained GPs worked as the first DR screening level of HUN from March 2015 to December 2023. They visualized NaIA-RD's proposals since the 1st of July 2020, when NaIA-RD was deployed. We have compared their screening decisions with the screening proposals of NaIA-RD, both before (retrospectively) and after (prospectively) they started using NaIA-RD. Clinical decisions and NaIA-RD grades were available as electronic health records. For the period prior to the deployment of NaIA-RD, we obtained the corresponding AI grades by calling NaIA-RD's API.

The mean patient age at the time of the eye study was 66.70 years (SD 12.65) before NaIA-RD was deployed, with 60.67\% males and 39.32\% females (sex assigned at birth). After NaIA-RD was implemented, the demographics remained very similar, with a mean age of 66.84 years (SD 13.33) and 60.48\% males and 39.51\% females.

\begin{figure}
    \centering
    \includegraphics[width=0.75\linewidth]{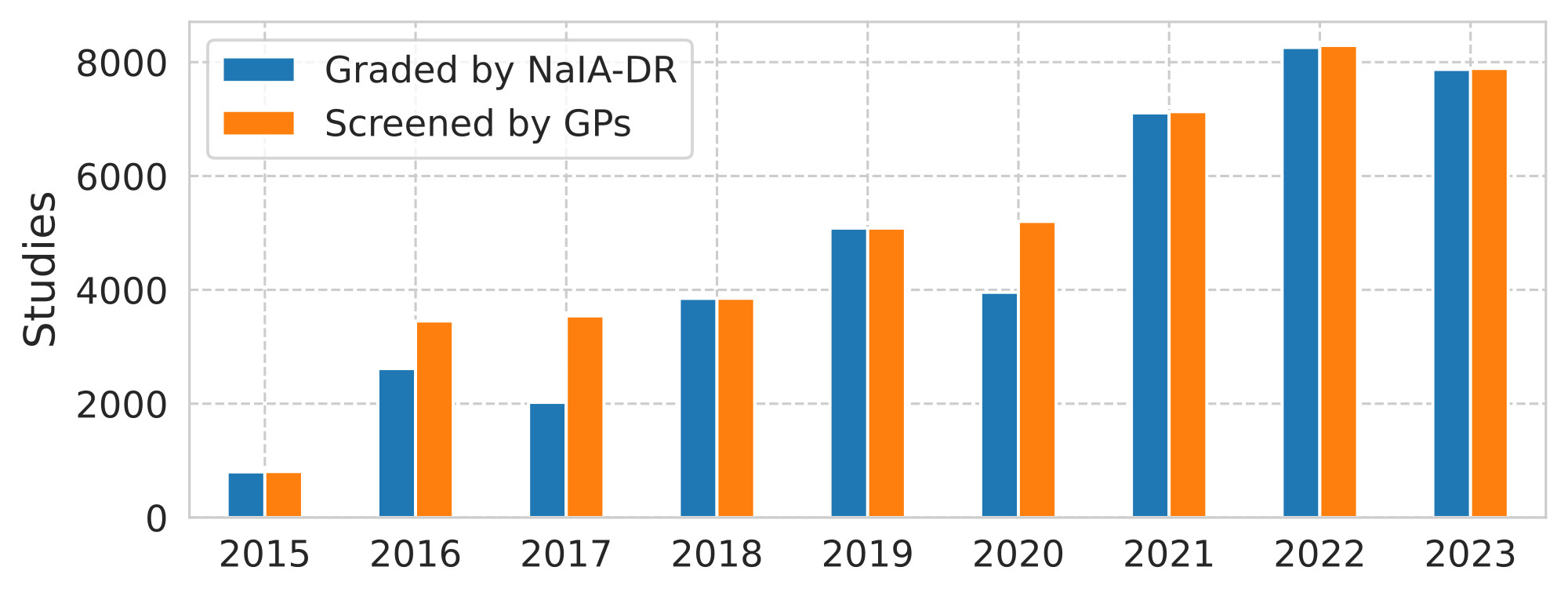}
    \caption{ Graded study quantity comparison: GPs vs NaIA-RD. }
    \label{fig:inferenced_study_qty}
\end{figure}

The volume of data that supports this study is illustrated in Figure \ref{fig:inferenced_study_qty}, with a histogram of the number of screened studies\footnote{We will define a study as an object consisting of digital retinographies, where each eye is represented by multiple fundus images. NaIA-RD grades each eye separately, and the global NaIA-RD screening proposal is referable if either eye is referable. Both the first and second screening levels made their decisions after visualizing both eyes (the entire study). See Section \ref{sec:dr-process} for more details.} per year. This number has increased since the start of the screening program in 2015, with the exception of 2023, when the number of screened patients slightly decreased. A NaIA-RD grade is available for the majority of the screened studies (81\% before deployment and more than 99\% after), but some NaIA-RD grades are missing due to image retrieval errors (3,685 grades in total). 

Using this data, we have examined the evolution of the DR screening program over time. We have analyzed the following aspects.

\vspace{6pt}
\noindent\textbf{Referral decisions versus NaIA-RD proposals}. To examine the relationship between decisions and AI grades in terms of volume, in Figure \ref{fig:referrals_vs_proposals} we compare the percentage of studies that GPs referred to the second screening level with the percentage that NaIA-RD graded as referable. Prior to its deployment in 2020, NaIA-RD would have proposed to refer slightly more studies than those referred by GPs (median\footnote{We used the median instead of the mean to minimize the influence of outliers.} difference of 2.74\%). However, this proportion increases in 2020 and afterwards (median difference between proposals and actual referrals of 9.16\%). We observe an anomaly in 2020 and 2021, when the proportion of referral proposals by NaIA-RD reaches 34.1\% and 35\%, respectively. Importantly, this anomaly only affects NaIA-RD: it did not lead to an increase in referral decisions by GPs, even though they were supported by NaIA-RD after July 2020. The anomaly ends in 2022, when the proportion of NaIA-RD referral proposals decreases to 24.6\% (2017 level), and reaches a minimum in 2023 (18\% studies). GPs' decisions also show this trend, as they refer fewer studies in 2022 and 2023 than ever before.

\begin{subfigure*}
    \centering
    \begin{minipage}[b]{.65\textwidth}
         \includegraphics[width=.99\linewidth]{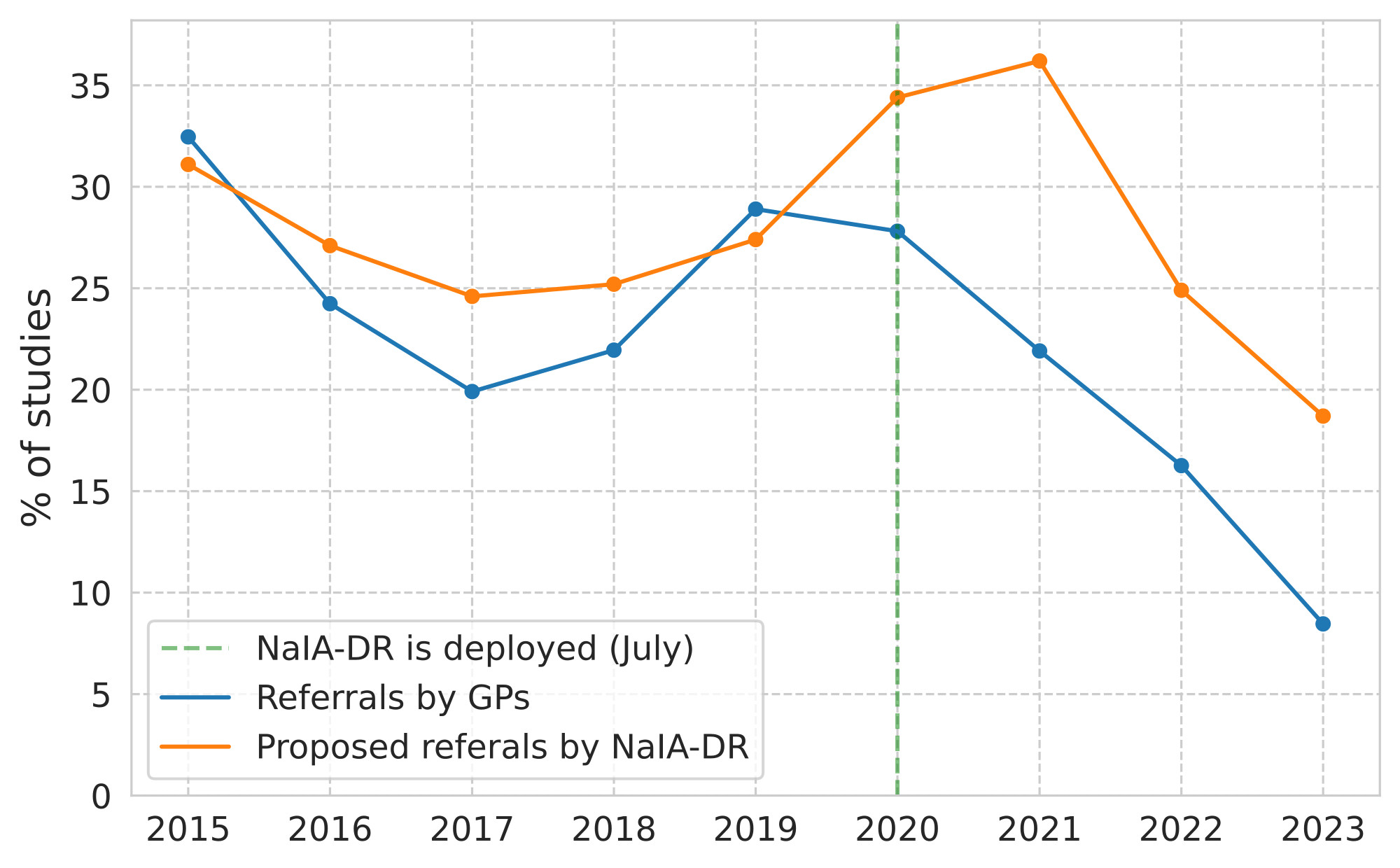}
        \caption{ Referral decisions versus NaIA-RD proposals. }
        \label{fig:referrals_vs_proposals}
    \end{minipage}
    \begin{minipage}[b]{.65\textwidth}
        \vspace{0.2cm}
        \includegraphics[width=.99\linewidth]{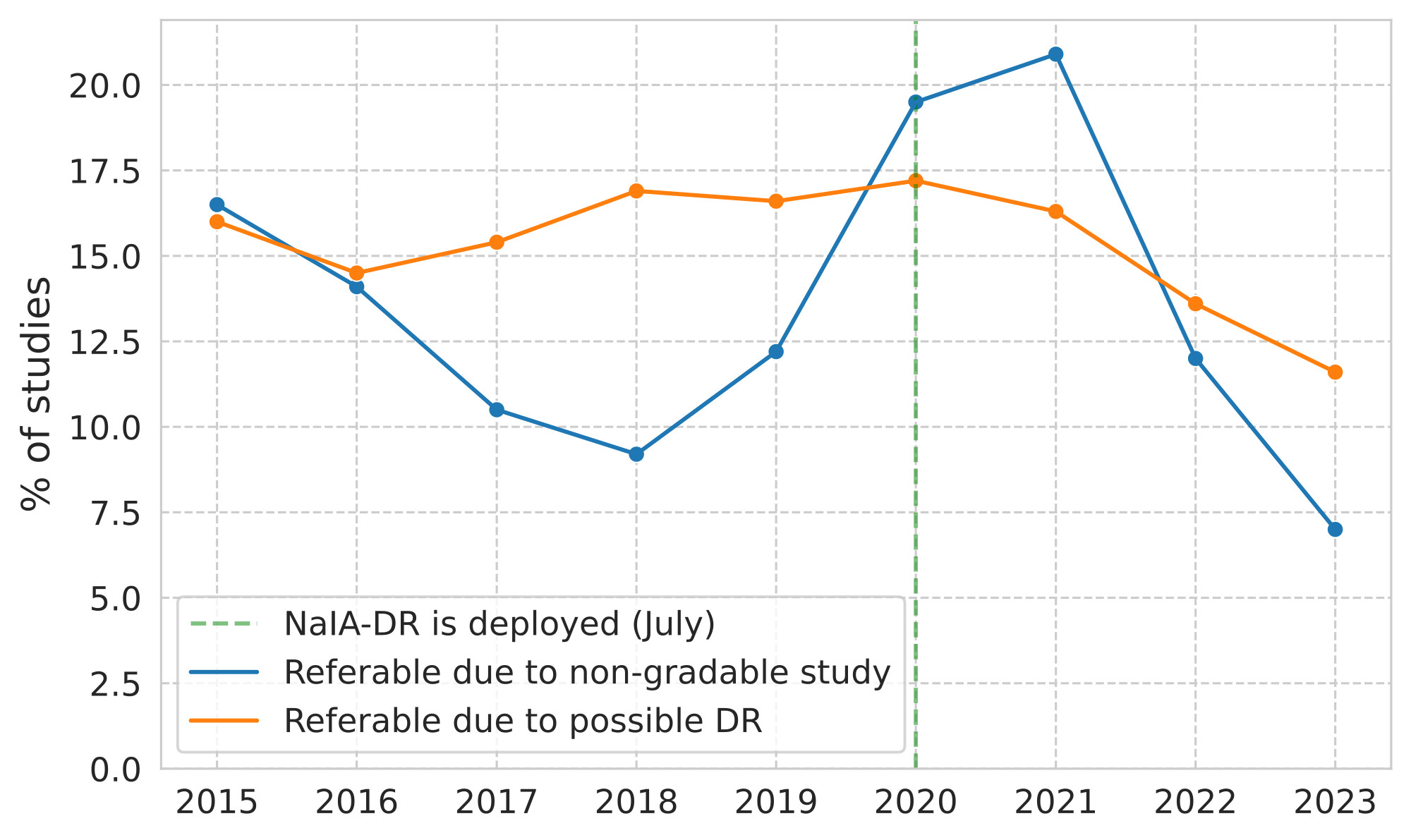}
    \caption{ Non-gradable versus referable DR (NaIA-RD outputs). }
    \label{fig:gradability_vs_severity}
    \end{minipage}
    \begin{minipage}[b]{.65\textwidth}
        \vspace{0.2cm}
        \includegraphics[width=0.99\linewidth]{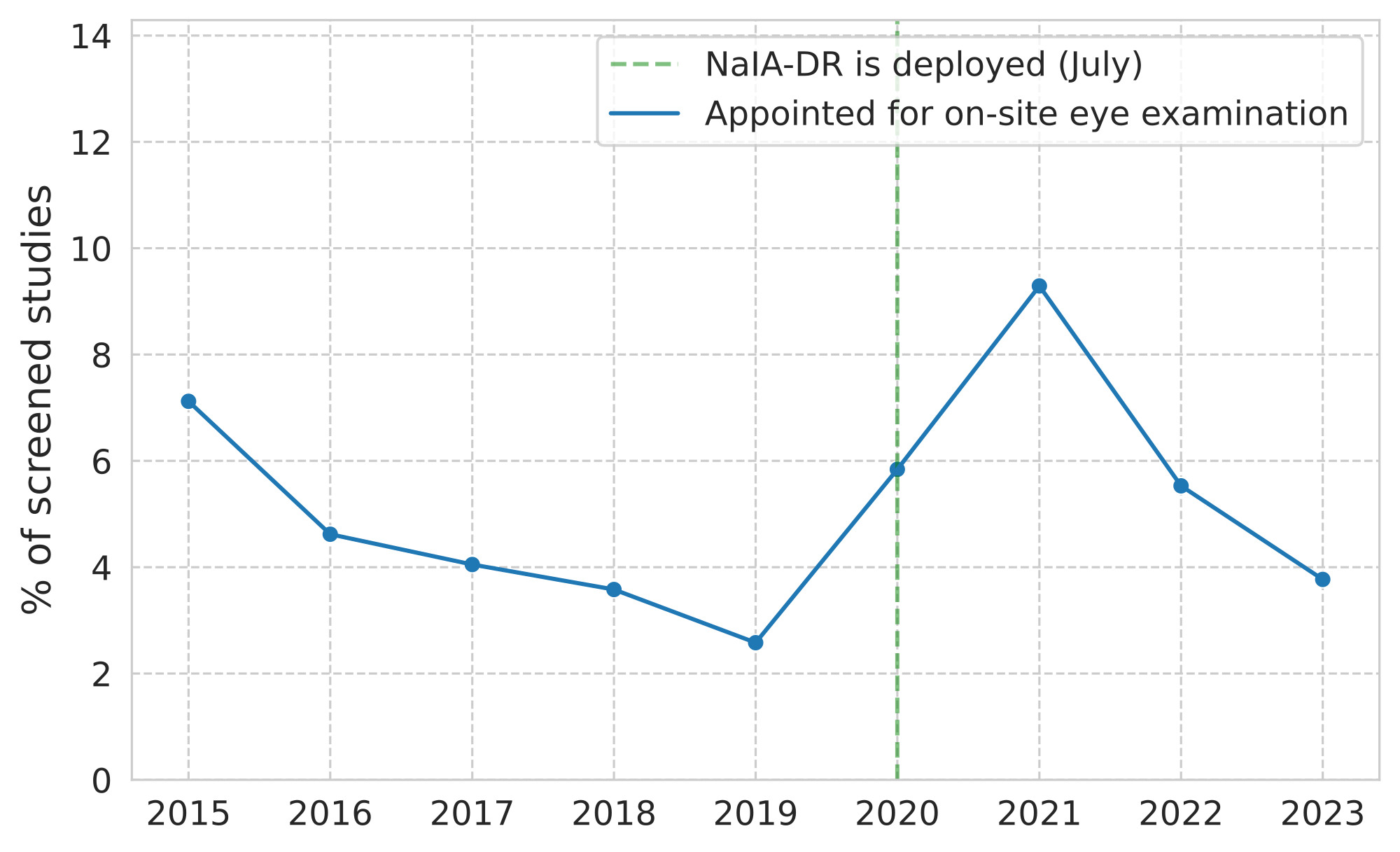}
        \caption{ Percentage of patients scheduled for on-site eye exam. }
        \label{fig:expert-cited-pct}
    \end{minipage}
    \setcounter{subfigure}{-1}
    \caption{ Evolution of the DR screening program of the HUN over time.}
    \label{fig:evolution-graphs}
\end{subfigure*}

\vspace{6pt}
\noindent\textbf{Non-gradable versus referable DR}. To try to explain the 2020-2021 anomaly, we have used the outputs of NaIA-RD to compare the evolution of the disease and the image gradability. Figure \ref{fig:gradability_vs_severity} shows  the percentage of referable studies due to possible DR and the percentage of referable studies due to non-gradability. While the proportion of possible DR studies has remained quite constant over time (between 11.6\% and 17.2\%), the proportion of non-gradable studies increases notably in 2020 and 2021 (from 12.2\% in 2019 to 20.9\% in 2021), while this proportion drops to historical minimums in 2023 (7\%).

\vspace{6pt}
\noindent\textbf{Patients requiring on-site eye examination}. To analyze the detection capability of the DR screening program, Figure \ref{fig:expert-cited-pct} shows the annual proportion of screened studies that resulted in an on-site eye examination. These eye exams were appointed by the second screening level, if the patient needed it, after the first-level screening GP (or the nurse due to high intraocular pressure) had referred the study. Patients left the DR screening program when this appointment occurred (see Section \ref{sec:dr-process} for more details). Note that the appointment proportion is calculated relative to the total number of studies screened. 

We observe that the appointment proportion decreases from 2015 to 2019, reaching a minimum of 2.58\%. However, coinciding with the deployment of NaIA-RD, this trend was broken in 2020, with a peak of 9.29\% in 2021, consistent with the observed non-gradability anomaly. In 2022 and 2023, despite the fact that the first screening level referred fewer studies than ever before (see Figure \ref{fig:referrals_vs_proposals}), the second screening level appointed more patients for eye examinations than in 2018-2019, achieving eye examination rates similar to the early years of the screening program. Specifically, the mean on-site eye examination proportion increased 1.5 times (from 3.08\% to 4.65\%), while 13.3\% fewer patients were referred (mean referral rate) in 2022-2023 compared to 2018-2019.

\vspace{6pt}
\noindent\textbf{Agreement between GPs and NaIA-RD}. To analyze the influence of NaIA-RD on clinical decisions, Figure \ref{fig:kappa_anual_GPs_global} shows the Cohen's kappa between first-level screening decisions and NaIA-RD outputs, a metric that measures the level of agreement between both. Note how the level of agreement more than doubles after the system was deployed (going from a kappa score of 0.2 in 2019 to a kappa score of 0.48 in 2023). To delve deeper, Figure \ref{fig:kappa_anual_GPs} shows the kappa score per first-level screening GP. We observe that some GPs had different screening criteria compared with others --they screened random studies from the same patient population, so we attribute these differences to screening criteria. Criteria differences occur particularly between two groups of GPs: ${1, 4}$ and ${2, 3}$. Interestingly, almost all GPs (except GP 4, who left the screening program in 2021 with few screened studies) clearly increased their agreement with NaIA-RD's proposals after its deployment. We would like to highlight the behavior of GP 1, who changed from an increasingly divergent agreement (reaching a minimum kappa score of 0.05 in 2019) to a much higher and constant level of agreement (kappa scores between 0.20-0.32) since the deployment of NaIA-RD. Anyway, GP 1's agreement was still minimal after this change.

\setcounter{figure}{9}
\begin{subfigure*}
    \setcounter{subfigure}{0}
    \centering
    \begin{minipage}[b]{.75\textwidth}
        \includegraphics[width=.99\textwidth]{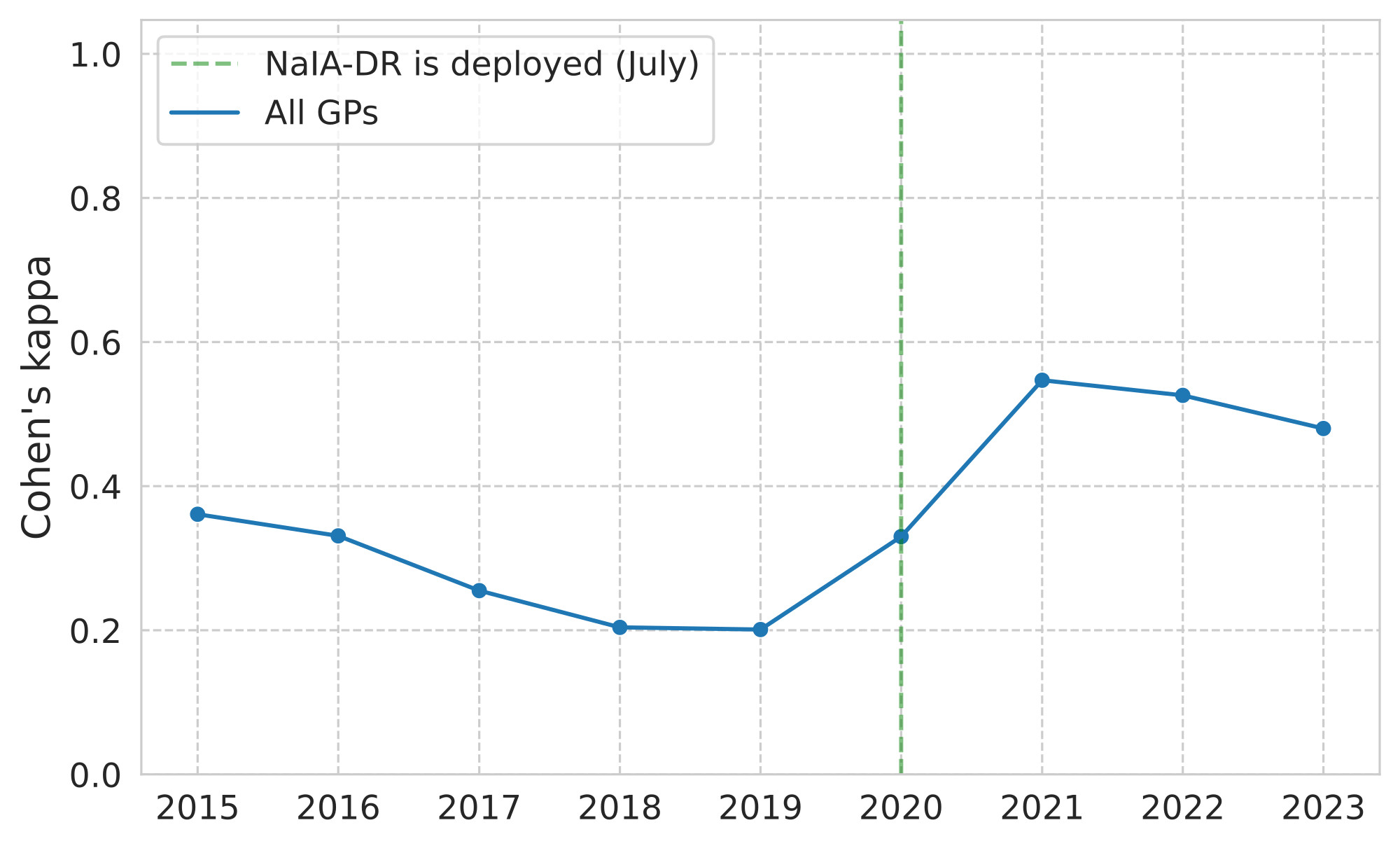}
        \caption{ Level of agreement between NaIA-RD and all GPs. }
        \label{fig:kappa_anual_GPs_global}
    \end{minipage}
    \begin{minipage}[b]{.75\textwidth}
        \vspace{0.5cm}
        \includegraphics[width=.99\textwidth]{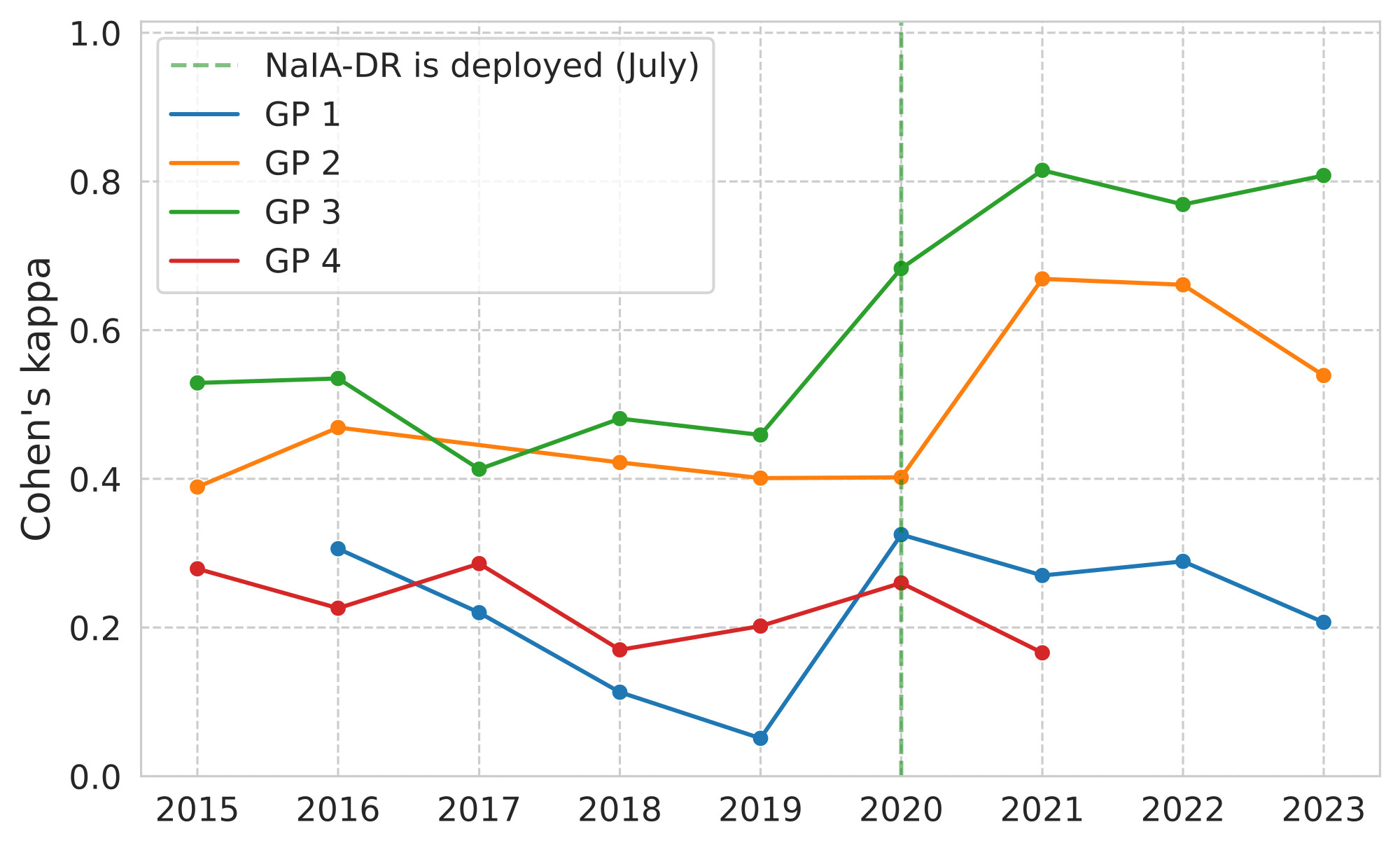}
        \caption{ Level of agreement between NaIA-RD and each GP. }
        \label{fig:kappa_anual_GPs}
    \end{minipage}
    \setcounter{subfigure}{-1}
    \caption{ Level of agreement between first-level screening GPs (GPs) and NaIA-RD's grades.}
    \label{fig:kappa_anual_prospective}
\end{subfigure*}

\vspace{6pt}
\noindent\textbf{NaIA-RD metrics compared to appointed on-site eye examinations per GP}. To further explore the impact of NaIA-RD, in Table \ref{prospective_GP_metrics_table} we compare NaIA-RD's post-deployment metrics (2020-2023) with the proportion of referrals and appointed on-site eye examinations for each first-level screening GP. We obtained NaIA-RD's positive agreement (PA), negative agreement (NA) and Cohen's kappa for each GP. PA and NA measure the proportion of NaIA-RD's referral and non-referral proposals that are agreed by the GP, respectively. Each row of this table takes the corresponding GP's decisions as the ground truth. The last three columns show the number of screened studies by each GP and two proportions: the rate of referred studies and the rate of studies that resulted in an on-site eye examination appointment. Both proportions are calculated relative to the total number of studies screened by each GP. 

\begin{table}
\centering
\begin{tabular}{ ccccrrg }
\toprule
 GP & PA\tnote{1} & NA\tnote{2} & Kappa & \multicolumn{1}{c}{Studies} & \multicolumn{1}{c}{Referred} & \thead{On-site \\ eye exams\tnote{3}} \\ 
\midrule
1 & 0.234 & 0.999 & 0.307 & 10,543 & 6.3\%   & 3.8\%  \\ 
2 & 0.612 & 0.980 & 0.664 & 6,902  & 16.7\% & 8.8\%  \\ 
3 & 0.866 & 0.946 & 0.812 & 5,089  & 29.2\% & 13.3\% \\ 
4 & 0.291 & 0.985 & 0.322 & 428    & 11.7\% & 6.1\%  \\ 
\bottomrule
\end{tabular}
    \scriptsize
    \begin{tablenotes}
        \item[1] PA: Positive Agreement. NaIA-RD proposes referral and the GP refers.
        \item[2] NA: Negative Agreement. NaIA-RD proposes non-referral and the GP does not refer.
        \item[3] On-site eye exams: Proportion of screened studies that resulted in an on-site eye examination appointment by the second screening level. A higher value represents a higher sensitivity of the GP for detecting target patients.
    \end{tablenotes}
\caption{\label{prospective_GP_metrics_table} NaIA-RD metrics compared to appointed on-site eye examinations per general practitioner (GP). All GPs took their decision after viewing NaIA-RD's proposal, once NaIA-RD was in use.
}
\end{table}

Data show that GPs who agreed more with NaIA-RD (Cohen's kappa) identified a higher proportion of patients that were appointed for an on-site eye examination. For example, GP 3, with a kappa of 0.812 and an appointment rate of 13.3\%, more than triples the appointment rate of GP 1, who shows a kappa of 0.307 and an appointment rate of 3.8\%. However, with more than 10,543 screened studies, GP 1 has screened almost as many studies as the rest of the GPs combined, which explains the weak kappa values shown in Figure \ref{fig:kappa_anual_GPs_global}. Finally, we observe high NAs for all GPs (0.946 or more), indicating that most of the NaIA-RD's non-referral proposals are followed. On the other hand, compared with GPs 2 and 3, the low PA of GPs 1 and 4 suggest a less sensitive referral criterion. This has resulted in lower referral rates, but also lower rates of patients appointed for on-site eye examinations. The global PA and NA are 0.493 and 0.981, respectively, with a global Cohen's kappa of 0.554.

\begin{figure}
\centering
\small
\begin{tikzpicture}[node distance=2cm, auto]
    \node[draw, align=center] (A) {Screened studies using\\ NaIA-RD between 2020-2021};
    \node[draw, below of=A, align=center] (B) { NaIA-RD proposed a referral\\but the GP did not refer};
    \node[draw, below of=B, align=center] (C) {The GP graded as\\Mild DR any of both eyes};
    \node[draw, below of=C] (D) {Random sample of 57.2\%};
    \node[draw, below of=D] (E) {False positive dataset};
    \draw[-latex] (A) -- node {9090 studies} (B);
    \draw[-latex] (B) -- node {1718 false positives} (C);
    \draw[-latex] (C) -- node {573 mild DRs} (D);
    \draw[-latex] (D) -- node {328 studies} (E);
\end{tikzpicture}
\caption{\label{err_analysis_dataset_method} Study selection process to create a false positive dataset.
}
\end{figure}
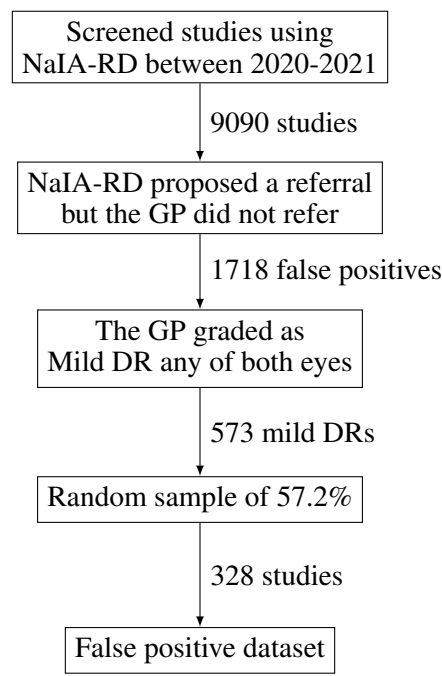

\vspace{6pt}
\noindent\textbf{Error analysis}. Due to the observed low PA, an expert ophthalmologist blindly labeled a sample of false positives of the 2020-2021 period, creating what we call a false positive dataset. In Figure \ref{err_analysis_dataset_method} we detail the study selection process we followed to create it. We wanted to review the most difficult and meaningful studies (potentially mild cases), so we selected studies that had been graded by the GP as mild non-proliferative DR. The ophthalmologist classified 93\% of the studies as referable (306 out of 328), matching NaIA-RD's proposals. Thus, GPs correctly contradicted NaIA-RD's referral proposals in only 7\% of the studies.

After reviewing a sample of NaIA-RD's false positives, we analyzed all false negatives from 2020-2023 period, which were only 180 studies (less than 1\% of the 22,962 total screened) --studies that NaIA-RD did not propose to refer but the GP referred. The second screening level had graded these false negatives according to the ICDR scale \cite{ICDR-2003}, with the following results: 83.33\% (150) were classified as "no apparent retinopathy", 6.67\% (12) as "mild non-proliferative DR", 1.11\% (2) as "moderate non-proliferative DR", 0.56\% (1) as "severe non-proliferative DR", 0\% (0) as "proliferative DR", and 10.66\% (29) as "not gradable". An ophthalmologist reviewed the one study that was graded as severe and could not find any DR signs on the images. Therefore, we attributed the grade to a data error.

\vspace{6pt}
\noindent\textbf{NaIA-RD as first screening level}. To assess whether NaIA-RD could reduce current human supervision, we analyzed what would have happened if it had performed the first level of screening autonomously --without GP supervision-- since 2020. During this period, 22,962 studies were screened in HUN, and 14.62\% were referred by GPs assisted by NaIA-RD. NaIA-RD would have referred 26.85\% of these, 1.84 times the current referred study quantity. However, the studies that NaIA-RD deemed non-referable would not have required human grading, which were the 73.15\%, and the referable studies would have been visualized only once (and not twice as in the current setup). Overall, it would have allowed a 4.27 times workload reduction (6,125 study visualizations instead of 26,318 current visualizations). 

\section{Discussion}
\label{sec:discussion}

We have found that NaIA-RD is a sensitive DR screening tool that has had a positive influence on the trained GPs who have screened patients from March 2015 to December 2023 at HUN. Our before-and-after study shows that these GPs are more capable of identifying patients who require on-site eye examinations since they started using NaIA-RD (see Figure \ref{fig:expert-cited-pct}). In fact, GPs were unable to identify a single case of sight-threatening DR missed by NaIA-RD since its deployment. This suggest that NaIA-RD could be safely and effectively used for autonomous, first-level DR screening at HUN.

In this section we will discuss the performance of NaIA-RD in laboratory settings (Section \ref{sec:performance}), its impact on the screening program of HUN (Section \ref{sec:impact}), the convenience of buying or developing such a system (Section \ref{sec:buy-or-develop}), and the limitations of this work (Section \ref{sec:limitations}).

\subsection{NaIA-RD's performance}
\label{sec:performance}
With a Cohen's kappa of 0.818, NaIA-RD shows a strong agreement with the majority label of three ophthalmologists according to our private Gold Standard, a metric comparable to the performance of a single ophthalmologist. A subset of this dataset also shows (retrospectively) that NaIA-RD is clearly more sensitive than a trained GP working in real world settings.

In terms of Gold Standard sensitivity and specificity, NaIA-RD achieved values of 92.5\% (95\% CI, 88.1-96.3\%) and 92.4\% (95\% CI, 89.5-94.9\%), respectively. These results clearly exceed the superiority endpoints of 85.0\% and 82.5\% proposed by Abràmof et al. in \cite{Abramoff2018USProspectiveIDX} for a DR screening medical device, as discussed by the authors with the FDA.

The goal of NaIA-RD is to improve the clinical pathway for DR screening at HUN. Therefore, it is not expected to be the most accurate DR screening tool that could be used in any other hospital. However, we have tested it on several public datasets dedicated to DR grading and retinal image quality assessment. Results suggest that NaIA-RD correctly assesses retinal images from diverse data distributions and patient populations.

With respect to DR assessment, we have tested NaIA-RD on EyePACS (Kaggle) \cite{KaggleEyePACSCompetition2016, EyePACS2009}, APTOS (Kaggle) \cite{APTOSKaggle2019}, Messidor-2 \cite{Messidor2018}, and IDRiD \cite{IDRIDDataset}, obtaining AUCs above 0.96 in all of them. These results do not seem to be far from the state of the art: For example, using an ensemble of five models trained on the Kaggle EyePACS training set, Papadopoulos et al. \cite{Papadopoulos2021} achieve an AUC of 0.961 on the public test set of Kaggle EyePACS, and 0.976 on Messidor-2.  Trying to replicate the well-known work of Gulshan et al. \cite{GulshanJama2016} using only public data,  Voets et al. \cite{VoetsReproduction2019} obtained an AUC of 0.951 on the Kaggle EyePACS public test set and 0.853 on Messidor-2 with an ensemble of ten Inception-V3 models. Voets et al. argued that they struggled to reproduce the original algorithm due to differences in the datasets used. Note that NaIA-RD was trained and tested on a subset of these datasets, so these metrics should be compared with caution. To test the generalization capabilities of the DR Classifier, we used OIA-DDR, which was completely excluded from training. An AUC of 0.957 and a sensitivity and specificity of 93\% and 84\% were obtained, but we could not find any other work with comparable results.

Regarding gradability assessment, we evaluated NaIA-RD's generalization capabilities using OIA-DDR \cite{OIA-DDR-2019} and EyeQ \cite{EyeQFu2019}. 
On OIA-DDR, the Gradability Classifier achieved perfect sensitivity, correctly identifying all non-gradable images, albeit with a 38\% false positive rate. However, it demonstrated higher specificity on EyeQ, with sensitivity and specificity values of 87\% and 86\%, respectively. Recall that EyeQ, unlike OIA-DDR, is a dataset specifically designed to assess retinal image quality.

While we explored alternative retinal image quality datasets like DeepDRiD \cite{DeepDRiDLiu2022} and DRIMDB \cite{DRIMDB2014}, they proved unsuitable for our purposes. DeepDRiD offered a potentially relevant overall quality class but suffered from inconsistent labeling, an issue acknowledged by its authors \cite{DeepDRiDLiu2022}. DRIMDB, on the other hand, provided cropped retinal images, incompatible with NaIA-RD.

These results show that NaIA-RD is more sensitive than specific for both DR and gradability assessment. This behavior ensures safety in a tool intended to be used as a first-level screening device. However, we also value its  performance as competitive: in a recent multicenter study \cite{Multicenter2021}, the best performing AI device had 80.47\% and 81.28\% of sensitivity and specificity on an external dataset (a performance comparable to human graders).

\subsection{NaIA-RD's impact}
\label{sec:impact}
We have presented a before-and-after study showing the impact of NaIA-RD on a real screening program. Results show that NaIA-RD has influenced the decisions of the first-level screening GPs, biasing them towards system's proposals. We interpret this bias as desirable, as GPs with a higher level of agreement with NaIA-RD were better at identifying patients who needed an on-site eye examination by a specialist (they were more sensitive). In fact, the rate of on-site eye examinations increased since the use of NaIA-RD, breaking a downward trend in the pre-NaIA-RD period (2015-2019 years in Figure \ref{fig:expert-cited-pct}).

Nevertheless, prospective results show that some first-level screening GPs such as 1 and 4 (for whom NaIA-RD shows low PAs in Table \ref{prospective_GP_metrics_table}) could have improved their sensitivity by more frequently following NaIA-RD's referral proposals. These GPs minimized the amount of studies that the second screening level received, but they were less able to identify patients who should have been referred compared with other GPs. The impact of these decisions is relevant, as GP 1 --who only followed the 23\% of referral proposals (PA 0.234)-- screened almost as many studies as the rest of the GPs combined. We discarded a malfunction of NaIA-RD with a false positive analysis dataset, in which an expert ophthalmologist labeled a sample of the contradicted referral proposals: the expert agreed with NaIA-RD in 93\% of the studies. Given that NaIA-RD was correct most of the time, we hypothesize that the observed variability in GP behavior may stem from a lack of standardized DR screening protocols and varying levels of trust in NaIA-RD and AI-based tools.

On the other hand, NaIA-RD's non-referral proposals were consistently accepted by GPs (high NAs on Table  \ref{prospective_GP_metrics_table}), with only 180 non-referral proposals  ($<1\%$ of the total screened) not followed. Among these, only 2 undetected moderate DR cases were found. NaIA-RD has not missed any worse DR case since its deployment (at least cases that GPs did refer). This high sensitivity raises the question of whether total human supervision is necessary. It appears that NaIA-RD could safely perform the first level of screening autonomously, similar to other commercial devices such as IDx-DR, EyeArt, Retmarker\cite{Ribeiro2014ProspectiveRetmarker} or SELENA+\cite{SingaporeSelena2023}. In this scenario, NaIA-RD would refer non-gradable and more than mild DR studies directly to the second screening level, reducing the study visualization workload by 4.27 times.

The before-and-after study also shows how GPs lowered their image quality requirements when there was an abnormal proportion of non-gradable fundus images in 2020 and 2021: they tried to find DR signs even in images that they had previously considered as non-gradable. We interpret this flexible behavior as a normal human adaptation to new circumstances (possibly related to COVID-19 pandemic), which avoided an excess of work in the second screening level. We believe that this illustrates how subjective fundus gradability assessment is, and how interesting a dynamic gradability adjustment feature may be for a DR screening algorithm. Also, this finding seems to justify the need for grading DR even when the image quality is poor, a feature that very few commercial devices offer.

\subsection{Buy or develop?}
\label{sec:buy-or-develop}
In Europe, any medical device is subject to the EU 2017/745 regulation \footnote{EU 2017/745: \url{https://eur-lex.europa.eu/legal-content/EN/TXT/HTML/?uri=CELEX:32017R0745&from=ES\#d1e1058-1-1}} and the specific laws of the country where it is developed (in Spain it is regulated by the Royal Decree 192/2023\footnote{ Royal Decree 192/2023: \url{https://www.boe.es/boe/dias/2023/03/22/pdfs/BOE-A-2023-7416.pdf}}). Any AI-based software that makes clinical recommendations must be approved by a Notified Body before it can be used for any purpose other than research. There are two ways to obtain this authorization: a CE mark, which allows the device to be marketed, or an in-house authorization, which is restricted to local use. The latter option is a valid regulatory pathway in Europe under certain circumstances. According to EU 2017/745, this allows healthcare institutions to manufacture medical devices for internal use if no equivalent device is commercially available.


We recommend considering a custom development if the healthcare institution has the required resources and does not find a commercial device that satisfies their requirements. However, the complexity of such development should not be underestimated: the access to target domain knowledge, data, software and IT infrastructure are essential. Dedicated AI and software development teams are needed, who must work closely with clinicians and the hospital. The regulatory work is also unavoidable, even in the case of an in-house authorization.

\subsection{Limitations}
\label{sec:limitations}
This work has some important limitations and knowledge gaps that need to be considered.

First, we have measured NaIA-RD's performance on the target data distribution only at eye level, not at image level. We could not measure how performance differed between macula-centered and optic disc-centered fundus images. The private Gold Standard does not provide a label per fundus image, but rather a label per eye. DR screening is performed at patient level, so the before-and-after study does not provide this separate measurement either.

Second, the comparison between NaIA-RD and a single ophthalmologist in the Gold Standard may be biased. We only had three ophthalmologists available for labeling, so when we compared each ophthalmologist's label to the other two, we used the grade of NaIA-RD to break ties.

Third, we could not test the performance of the Gradability Classifier using an adequate private dataset. Due to the low prevalence of bad quality images, the private Gold Standard is not appropriate for measuring the sensitivity of the eye fundus Gradability Classifier. Therefore, we tested it using external datasets. 

Fourth, we did not test NaIA-RD's performance in population subgroups, nor did we perform a systematic bias analysis. Although the overall results are promising, such an analysis should be performed to discard any bias due to imaging site, sex, gender, age or pre-existing patient conditions.

Fifth, with respect to the generalization ability of NaIA-RD, it should be noted that the DR Classifier was trained on a subset of the most popular public datasets (see Figure \ref{fig:data-sets-in-train}). Therefore, its metrics may be biased when evaluating these datasets. On the other hand, the metrics we obtained evaluating OIA-DDR and EyeQ (which were not used in training) are promising, but further testing is needed to assess the generalization capabilities of NaIA-RD's models.

Sixth, our comparison of commercial devices is based solely on public information available prior to 2024 --we did not contact any vendors. 
The lack of a detailed public database of CE marked devices in Europe did not facilitate our research \cite{Approval2021}. 
We therefore advise against using our product overview as a purchasing guide.

Seventh, due to the lack of a control group in our before-and-after study, we cannot conclusively attribute the observed improvements in the DR screening program solely to NaIA-RD. It is possible that other factors influenced the change in GP behavior. A different type of clinical study, such as a randomized controlled trial, may have provided stronger evidence.

Eighth, we did not compare NaIA-RD with other commercial devices in terms of accuracy. This comparison was beyond the scope of this work: our goal was to provide a capable (not the most accurate) AI tool for DR screening at HUN (not other hospitals) that met the requirements that commercial devices did not.

Ninth, due to copyright and privacy concerns, this work does not publish any code or dataset. Although this fact limits its reproducibility, we believe that other researchers can use this work to develop their own in-house AI medical device --given that they have their own data.

This work also leaves some important questions unanswered. First, we weren't able to assess patient benefit: Did patients have better DR outcomes after NaIA-RD was in use? A randomized controlled trial seems necessary to answer this key question. In addition, we could not clarify why there are some first-level screening GPs who contradict the system much more than others.

\subsection{Conclusion}
\label{sec:conclusion}
We have presented an AI-powered medical device, called NaIA-RD, customized to the needs of a hospital (HUN). We have detailed the entire development process with a level of detail that should allow other researchers to follow our approach using their own data. We have proposed a novel system design that combines DR grading and retinal image quality in a single, flexible AI device. We have measured its performance on private and public datasets, achieving competitive metrics, and assessed its real-world impact with a before-and-after study that is unprecedented in the literature.

We have found that NaIA-RD increased the sensitivity of the GPs who conformed the first screening level: these clinicians were influenced by the AI tool, increasing the proportion of patients who were scheduled for eye examination by the second screening level --which were the target patients to be detected. However, GPs showed a very heterogeneous screening criteria, and NaIA-RD did not influence them equally.

We have also observed that GPs adapted their DR screening decisions to seasonal anomalies, such as a sudden deterioration in image quality. When this occurred, GPs lowered their standards for fundus gradability. This clinical adaptability implies that there is a friction between best practices, which are strictly followed by AI tools, and their real-world applicability. To address this issue, it appears that DR screening tools should be adjustable to local needs. Calibrated decision thresholds should help achieving this to some extent, as long as AI tools separate disease from image quality assessment.

In spite of their lack of context awareness, we conclude that AI tools can improve and homogenize DR screening, while reducing the burden of manual grading: NaIA-RD could safely be used to perform first level screening autonomously, reducing the workload by up to 4.27 times, at the expense of receiving more studies in the second screening level (1.84 times more). Therefore, our future work will be directed towards reducing the level of supervision of NaIA-RD, accompanied by new features to facilitate the task of the second screening level. In our opinion, an AI-generated ICDR \cite{ICDR-2003} grade should be useful to automate clinical reports in a supervised manner.

We believe NaIA-RD exemplifies how an AI medical device can drive long-term clinical improvements. Its seamless integration into the clinical workflow appears to be a key factor in its success --potentially as important as its accuracy. Healthcare institutions should be aware that commercial AI tools are typically designed to work with specific patient populations, cameras, and imaging protocols, offering limited options for adaptation and customization. This can lead to challenges when incorporating them into existing clinical workflows. In such cases, developing an in-house solution may be a good alternative, provided the institution has the necessary resources.

\section*{Ethics statement}
This work was approved by the Health Care Ethics Committee of the Hospital University of Navarre. All clinical records and DICOM studies were anonymized before being used, and no individual patient clinical history was accessed at any time. Therefore, no patient consent was needed for this work \cite{anonym-data}. NaIA-RD is a non-commercial, in-house software developed by, and used exclusively by HUN for research purposes. As such, it does not require a CE mark (as stipulated by EU Regulation 2017/745). All the public datasets used for NaIA-RD development and validation explicitly authorize their usage under these conditions. All the tools used for NaIA-RD development are open source.

\section*{Statement on conflicts of interest}
Mikel Galar and Jose Andonegui declare no competing financial interests or personal relationships; all other authors declare an employment relationship with the Government of Navarre and a related financial support.

\section*{CRediT authorship contribution statement}
\textbf{Imanol Pinto:} Conceptualization, Investigation, Formal analysis, Software, Data Curation, Writing - Original Draft. \textbf{Álvaro Olazarán:} Conceptualization, Investigation, Formal analysis, Software, Data Curation, Writing - Review \& Editing. \textbf{David Jurío:} Conceptualization, Investigation, Formal analysis, Software, Data Curation, Writing - Review \& Editing. \textbf{Borja de la Osa:} Conceptualization, Investigation, Formal analysis, Writing - Review \& Editing. \textbf{Miguel Sainz:} Writing - Review \& Editing. \textbf{Aritz Oscoz:} Investigation, Project administration, Writing - Review \& Editing. \textbf{Jerónimo Ballaz:} Project administration, Writing - Review \& Editing. \textbf{Mikel Galar:} Conceptualization, Methodology, Supervision, Writing - Review \& Editing. \textbf{Javier Gorricho:} Supervision, Resources, Writing - Review \& Editing. \textbf{José Andonegui:} Conceptualization, Supervision, Resources, Validation, Data Curation, Writing - Review \& Editing.

\section*{Acknowledgments}
NaIA-RD would not exist without the invaluable work of many people. From the University Hospital of Navarre (HUN) and the Navarre Public Health Service (SNS-O), we would like to express our special thanks to Javier Turumbay, Elena Manso, Alejandro Dávila, Marcos Mozo, Fermín Bruque, María Jesús Esparza and Begoña Martínez. From the General Directorate of Telecommunications and Digitalization (DGTD), we would like to highlight the work of Jokin Sanz, Adrián Errea, Carlos Romero, Ainhoa Pagola, the Electronic Health System Team (HCI), the Digital Imaging Team, the Infrastructure Team and the Architecture Group (GAT).

\section*{Funding}
NaIA-RD has been fully funded by the Government of Navarre.

\section*{Data Availability Statement}
Private datasets of NaIA-RD can be made available after reasonable request.

\bibliographystyle{Frontiers-Vancouver} 
\bibliography{test}



\end{document}


\onecolumn
\firstpage{1}

\title[Supplementary Material]{{\helveticaitalic{Supplementary Material}}}

\maketitle

\section{Supplementary Data}

Supplementary Material should be uploaded separately on submission. Please include any supplementary data, figures and/or tables. All supplementary files are deposited to FigShare for permanent storage and receive a DOI.

Supplementary material is not typeset so please ensure that all information is clearly presented, the appropriate caption is included in the file and not in the manuscript, and that the style conforms to the rest of the article. To avoid discrepancies between the published article and the supplementary material, please do not add the title, author list, affiliations or correspondence in the supplementary files.

\section{Supplementary Tables and Figures}

For more information on Supplementary Material and for details on the different file types accepted, please see \href{http://home.frontiersin.org/about/author-guidelines#SupplementaryMaterial}{the Supplementary Material section} of the Author Guidelines.

Figures, tables, and images will be published under a Creative Commons CC-BY licence and permission must be obtained for use of copyrighted material from other sources (including re-published/adapted/modified/partial figures and images from the internet). It is the responsibility of the authors to acquire the licenses, to follow any citation instructions requested by third-party rights holders, and cover any supplementary charges.